\documentclass{emulateapj}
 \usepackage{psfrag}
 \usepackage{rotating}
 \newcommand{\arc}{^{\prime\prime}}
 \shorttitle{Mass Loss in LMC Cepheids II.}
 \shortauthors{Neilson et al.}
  
\begin{document}
  
 \title{Testing Mass Loss in Large Magellanic Cloud Cepheids using Infrared and Optical Observations II. Predictions and Tests of the OGLE-III Fundamental-Mode Cepheids}
\author{Hilding R. Neilson\altaffilmark{1,2}}
\altaffiltext{1}{Argelander Institute for Astronomy, University of Bonn, Auf dem H\"{u}gel 71, Bonn, 53121, Germany; hneilson@astro.uni-bonn.de}
\altaffiltext{2}{Department of Astronomy \& Astrophysics, University of Toronto, 50 St. George Street, Toronto, ON,  M5S 3H4, Canada}
\author{Chow-Choong Ngeow\altaffilmark{3}}
\altaffiltext{3}{Graduate Institute of Astronomy, National Central University, Jhongli City, 32001 Taiwan (R.O.C)}
\author{Shashi M. Kanbur\altaffilmark{4}}
\altaffiltext{4}{Department of Physics, State University of New York at Oswego, Oswego, NY 13126, USA} 
\author{John B. Lester\altaffilmark{2,5}}
\altaffiltext{5}{University of Toronto Mississauga, Mississauga, ON, L5L 1C6, Canada}
\begin{abstract}
In this article, we test the hypothesis that Cepheids have infrared excesses due to mass loss.  We fit a model using the mass-loss rate and the stellar radius as free parameters to optical observations from the OGLE-III survey and infrared observations from the 2MASS and SAGE data sets.  The sample of Cepheids have predicted minimum mass-loss rates ranging from zero to $10^{-8}M_\odot$ $yr^{-1}$, where the rates depend on the chosen dust properties.  We use the predicted radii to compute the Period-Radius relation for LMC Cepheids, and to estimate the uncertainty caused by the presence of infrared excess for determining angular diameters with the infrared surface brightness technique.  Finally, we calculate the linear and non-linear Period-Luminosity (P-L) relations for the LMC Cepheids at $VIJHK$ + IRAC  wavelengths and we find that the P-L relations are consistent with being non-linear at infrared wavelengths, contrary to previous results.  
\end{abstract}

\section{Introduction}
Classical Cepheids are powerful standard candles because they follow the  Period-Luminosity (P-L)  relation or Leavitt Law \citep{Leavitt1908}. The P-L relation has been used as a tool for Galactic \citep{Feast1997b}, extragalactic \citep[and references therein]{Gieren2009} and cosmological studies \citep{Freedman2001, Sandage1998}.  The LMC Cepheids have been central for deriving these relations, initially using optical data \citep{Udalski1999a,Sandage2004,Kanbur2003, Kanbur2006, Fouque2007}, then near-infrared observations \citep{Groenewegen2000, Persson2004, Fouque2007}, and recently using infrared observations \citep{Freedman2008, Ngeow2008a}.  The infrared P-L relations were determined by matching sources from the SAGE (Surveying the Agents of a Galaxy's Evolution) archival data \citep{Meixner2006} with the known Cepheids.  

\cite{Ngeow2008b} computed the infrared P-L relations for a larger sample of 1848 Cepheids from the OGLE-III Database \citep{Soszynski2008} by matching those sources with the two published epochs of SAGE data, and also averaging the two epochs of infrared fluxes to bring the fluxes closer to the mean flux. They tested the P-L relations for non-linearity, and found that the IR P-L relations are linear, while the relations at wavelengths shorter than $K$-band are non-linear.  However, there are a number of Cepheids with infrared fluxes that are $>3$ standard deviations brighter than predicted by the P-L relations. Similar results were found using AKARI N-band data \citep{Ngeow2010}. It is important to understand the nature of this excess infrared brightness because the intrinsic infrared flux of Cepheids is less metallicity dependent than optical P-L relations \citep{Freedman2008}, which makes the IR P-L relations powerful tools for extragalactic studies.  The James Webb Space Telescope will be able to observe Cepheids in distant galaxies, making it possible to use the IR P-L relations to determine distances, but the unknown source of infrared excess will increase the uncertainty of the results.

One hypothesis for the source of the infrared excess is the existence of a circumstellar envelope (CSE) of dust around the LMC Cepheids, analogous to the CSEs observed around Galactic Cepheids \citep{Kervella2006, Merand2006, Merand2007}.  Multi-wavelength observations of RS Puppis and $l$ Car confirm the presence of CSEs around these Cepheids \citep{Kervella2009}. There is more evidence of infrared excess in Galactic Cepheids using IRAS observations \citep{McAlary1986, Deasy1988} and Spitzer observations \citep{Marengo2009}. On the other hand, \cite{Marengo2009b} argued that there is no evidence from IR excess from warm dust based on Spitzer observations and instead was due to carbon monoxide emission.  The Cepheid RS Pup is also associated with a nebula \citep{Havlen1972, Kervella2008}. 

\cite{Kervella2006} speculated that CSEs surrounding Galactic Cepheids were caused by mass loss that can leads to an infrared excess. \cite{Neilson2008a} proposed  a driving mechanism for this wind that is a combination of radiative acceleration, pulsation and shocks in the atmosphere of the Cepheid. This proposed driving mechanism  was tested by developing an analytic mass-loss model that was used to calculate rates of mass loss for a sample of Galactic Cepheids, with values between $10^{-10}$ to $10^{-7}M_\odot/yr$ being found.  This model predicts the fluxes of CSEs around the Cepheids that agree well with those observed.   \cite{Neilson2008b} applied this mass-loss model to theoretical models of Galactic, LMC, and Small Magellanic Cloud Cepheids and found that Cepheids in all three galaxies have significant mass-loss rates and infrared excesses.   There is other evidence for Cepheid mass loss in addition to infrared excess; \cite{Nardetto2008b} observed H$\alpha$ line profiles of Galactic Cepheids with the purpose of detecting a hydrogen circumstellar medium.  They found that the H$\alpha$ line profiles are asymmetric and blue-shifted in long period ($P>10$ day) Cepheids, possibly caused by stellar wind.   

While it is important to characterize infrared excess of Cepheids in order to use them for various applications, it is just as important to understand how mass loss affects the structure and evolution of Cepheids.  Mass loss is a potential solution to the mass discrepancy of Cepheids, which is the difference between Cepheid masses estimated  using stellar evolution calculations and estimates using stellar pulsation calculations \citep{Cox1980}.  Currently, stellar pulsation calculations predict masses about $10$-$20\%$ smaller than stellar evolution models \citep{Keller2006, Keller2008}. Measurements of dynamic masses of Cepheids in binary systems tend to agree with pulsation calculations \citep{Evans2006, Evans2008}, suggesting the discrepancy is in the stellar evolution modeling.  If Cepheids lose an average of $10^{-8}$ to $10^{-7}M_\odot/yr$, then mass loss could solve the mass discrepancy.

Mass loss is hypothesized to create optically thin circumstellar dust shells that cause infrared excess in Cepheids but do not affect the visual extinction of the Cepheids.  This excess should be seen in infrared observations of the LMC from the SAGE survey \citep{Meixner2006} for which there are currently two epochs of observations.  In fact, infrared excess has been detected in a number of evolved AGB stars in the LMC \citep{Vijh2008, Groenewegen2009}.    In our earlier paper testing mass loss in LMC Cepheids, we found that mass loss may be important, with about $10\%$ of OGLE-II Cepheids showing statistically significant evidence for dust shells. However,  the uncertainty of mean infrared flux of the Cepheids is a limiting factor.  The predicted gas mass-loss rates of the LMC Cepheids in that sample range from $10^{-11}$ to $10^{-8}M_\odot/yr$ assuming a dust-to-gas ratio of $1/250$.

Mass-loss rates in this range affect both the zero-point and the slope of the infrared P-L relations.  When the infrared excess was removed from the OGLE-II sample, \cite{Neilson2008c} found that the scatter of the data was reduced. We also found that the $3.6$ to $5.8$ $ \mu m$ relations are consistent with being non-linear while the $8.0$ $\mu m$ relation are linear, differing from the observed relations \citep{Ngeow2008a}.  The infrared excess caused by mass loss appeared to have a greater effect at shorter periods than at longer periods.  However, the infrared stellar luminosity is an increasing function of period, making the infrared excess is less noticeable for long-period Cepheids, even for larger mass-loss rates  ($\approx 10^{-8}~M_\odot/yr$).

These results suggest that the near-infrared fluxes, $JHK$, are potentially affected by mass loss.  This would mean distance estimates using the Infrared Surface Brightness (IRSB) technique have uncertainty due to mass loss and infrared excess.  The infrared surface brightness technique uses an empirical fit of the dependence of surface brightness on $(V-K)_0$ to determine the mean value and amplitude of a Cepheid's angular diameter  \citep{Gieren1999}. This angular diameter information is then converted to a radius and a distance by using radial velocity observations to compute the amplitude of the radius variation.  This technique has been shown to be robust, but infrared excess from mass loss suggests that the color $(V-K)_0$ may be overestimated.

The objective of this paper is to analyze fundamental-mode Cepheids in the LMC that have stellar fluxes from the OGLE-III survey and infrared data from the 2MASS and from two epochs of the SAGE surveys, and test if the Cepheids have infrared excess.  We determine mass-loss rates from infrared excesses determined from a correlation of these data sets by assuming a mass-loss model.  In the next section we describe the method for testing the model.  In section 3 we fit the model assuming zero mass loss, and the results of fitting the mass-loss model are given in section 4.  In section 5, we test how the removal of the $5.8$ and $8.0$ $\mu m$ data affect the mass-loss model as well as how the unknown pulsation phase of the IR fluxes affects the predicted mass-loss rates.  The statistical comparison of the two models and the uncertainty of the fits due to the unknown pulsation amplitudes of the IR fluxes is given in section 5.  Using the results of the models, we calculate the Period-Radius relation (section 6), and test the uncertainty of the IRSB technique due to mass loss (section 7). In Section 8 we compute linear IR P-L relations from the data and compare the predictions with the observed results of \cite{Ngeow2008b}, \cite{Freedman2008}, and \cite{Madore2008} in Section 9.  In Section 10, we test the P-L relations for non-linearity.

\section{Method for Determining Mass-Loss Rates and Radii of Cepheids}
To model the mass-loss rates of LMC Cepheids, we employ the sample of Cepheids from the OGLE-III survey with $V$ and $I$-band data. To obtain $JHK$ and IRAC observations, this sample is correlated with the 2MASS and SAGE surveys. These data, from \cite{Ngeow2008b}, have fluxes at up to nine wavelengths that have been corrected for extinction at all wavelengths.   For Cepheids with measurements in both epochs of the SAGE data, we follow the practice \cite{Ngeow2008b} of averaging the IRAC fluxes to reduce the uncertainty  due to pulsation phase and amplitude of brightness variation.

\begin{figure}[t]
\begin{center}
	\epsscale{1.2}
		\plotone{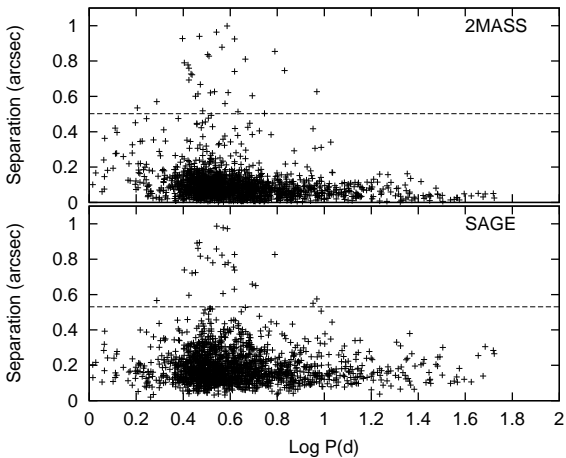}
\caption{The separation of the sources from the 2MASS (top) and SAGE (bottom) surveys matched with the fundamental mode Cepheids from the OGLE-III survey. The dashed lines represent the separation that is the average plus $3\sigma$ deviation. }
	\label{f1}
	\end{center}
\end{figure}

Our operating assumption is that infrared excess is due to a dust shell formed at a large distance from the Cepheid that is optically thin at visible wavelengths, but other explanations are possible. For example, an infrared excess could be caused by blending of stars  or by mistaken associations. In our previous work, false associations were possible because the average separation of SAGE and OGLE-II sources was about $0.77\arc$, with some Cepheids having separations as large as $1$ to $2\arc$.  Note that the uncertainty of the coordinate systems of the surveys are smaller than the separation: the average uncertainty of the OGLE-II coordinates are  about $0.1-0.2\arc$ \citep{Szymanski2005}. Using the OGLE-III data, the mean separation with the SAGE sources is about $0.2\arc$ , and the mean separation with 2MASS sources is about $0.1\arc$. However, the calibration error of the OGLE-III coordinates is only $0.06\arc$ (Udalski et al. 2008).  The separation of the sources is shown in  Figure \ref{f1}.  To guard against false associations, Cepheids in the sample with separations greater than the average plus $3\sigma$ deviation have been removed.

We model the mass-loss rate and radius of each Cepheid in the sample using the same method as in \cite{Neilson2008c} with several improvements. The observed luminosity at each wavelength is assumed to be the sum of the stellar luminosity and the contribution due to the dust shell.  Furthermore, we assume that the dust is composed of silicates instead of carbon because most Cepheids are typically not carbon rich.  Assuming silicon instead of carbon decreases the predicted mass-loss rate by a factor of $2$-$3$. The stellar luminosity is 
\begin{equation}\label{e1}
L_\nu (\rm{Star}) = 4\pi R_*^2 \pi B_\nu(T_{\rm{eff}}),
\end{equation}
and the contribution due to a dust shell is
\begin{eqnarray}\label{e2}
\nonumber &&L_\nu(\rm{Shell}) = 3\pi\frac{<a^2>}{<a^3>} \frac{\dot{M}_d}{\bar{\rho}v_d}Q^A_\nu \times \\
&&\mbox{\hspace{1cm}}\int_{R_c}^\infty B_\nu(T_d)[1-W(r)]dr.
\end{eqnarray}
This luminosity is different by a factor of $4\pi^2$ from the luminosity used in the first article and is the correct derivation of the relation; this is an error in the first article and the resultant mass-loss rates are a factor $4\pi^2$ too large but the other results remain the same.  In our first article, \citep{Neilson2008c}, we checked how well a blackbody represents the Cepheid flux in the $V$- and $I$-bands.  The difference is less than a few tenths of a magnitude.  The dust shell luminosity is dependent on the ratio of the dust particle's surface area to volume, $<a^2>/<a^3>$, the mean mass density of dust, $\bar{\rho}$, the mass-loss rate, $\dot{M}_d$, and the velocity of dust, $v_d$.  The variable $Q^A_\nu$ is the absorption efficiency, which is treated more precisely in this work.  Before, we assumed the efficiency was $\approx 2$ and here we treat it as a function of wavelength, $Q_\nu^A \approx 2\pi <a>/\lambda$,  an approximate relation that is reasonable for $\lambda < 10 \mu m$.  The dilution factor, $W(r)$ is a measure of the radiative equilibrium of the gas and dust with stellar radiation. Previously, we assumed that both the dust and gas had the same dilution factor. \cite{Ivezic1997} argued that the temperature of the dust varies as $\propto r^{-1/2}$, but the temperature of the gas goes as $\propto r^{-2/5}$.  In this work, we use the representation of \cite{Ivezic1997}.  The dust is assumed to form at the condensation temperature of $1200$ $K$, which leads to a dust-condensation radius,  $R_d \approx (R_*/2) (T_{\rm{eff}}/1200~K)^{2}$.  The dust particles are assumed to range in size from $a = 0.005$ to $0.25$ $\mu m$ with a mean density of $3.7$ $g/cm^3$, and to follow the \cite{Mathis1977} distribution.  The dust travels at a velocity similar to the escape velocity  of order $100$ $km/s$.  These assumptions are discussed in earlier works \citep{Neilson2008a, Neilson2008c}.

The effective temperature of the Cepheids is given by the color $(V-I)_0$, corrected both for extinction \citep[see][for details]{Ngeow2008b} and for the pulsation period from \cite{Beaulieu2001}.  This leaves the Cepheid radius, $R_*$, and the dust mass-loss rate as free parameters, and the dust mass-loss rate is converted to gas mass-loss rate by assuming a dust-to-gas ratio of $1:250$, which is the canonical Galactic value scaled by the relative metallicity of the LMC.  All of the following figures and discussion regarding the mass-loss rates refer to the gas mass-loss rate unless specified otherwise. Therefore we have two free parameters to fit observed magnitudes at up to nine wavelengths for each Cepheid.  We fit the model by calculating the minimum $\chi^2$ value.  This requires assuming both a distance to each Cepheid and an uncertainty of the flux due to pulsation phase. We assume a distance modulus of $18.5$ magnitudes to the LMC with an uncertainty of the distance being $\pm 0.1$ magnitudes due to the thickness of the LMC \citep[see][for details]{Neilson2008c}.  A final improvement is to assign an uncertainty to the flux due to  pulsation.  Following \cite{Ngeow2008a}, we set the uncertainty of the infrared flux at $1/3$ the $I$-band amplitude, instead of  assuming a constant value of the uncertainty. 

The approach of the following two sections is first to establish a  basis of comparison by determining the Cepheid radii assuming zero mass loss.  Then we calculated the mass-loss rates of the  Cepheids and use the $F$-test \citep[][and references therein]{Kanbur2004a, Ngeow2008b},
 to determine the quality of the fit for the mass loss and the zero mass loss cases, where the value of $F$ is 
\begin{equation}\label{e3}
F = \frac{\sigma_{1}^2(N-1) - \sigma_2^2(N-2)}{\sigma_2^2},
\end{equation}
 where $\sigma_1^2$ and $\sigma_2^2$ are the uncertainties of the fit using the Cepheid radius only and then using both the Cepheid radius and mass-loss rate, respectively, and $N$ is the number of data points fit.  Therefore, we may calculate the mass-loss rates of the LMC Cepheids and test the validity of the model.  In the fit, we do not necessarily have information for all nine wavebands, therefore Cepheids are rejected if either the $V$ or $I$ is not present or if two or more of the IRAC bands are not present or if fewer than five of the bands are present.  Out of the $1848$ fundamental mode Cepheids in the OGLE-III survey, this leaves $1398$ Cepheids from SAGE epoch 1 observations, $1387$ from epoch 2 and $1552$ Cepheids from the average of the two epochs.
 \begin{figure*}[t]
\begin{center}
	\epsscale{1.15}
		\plottwo{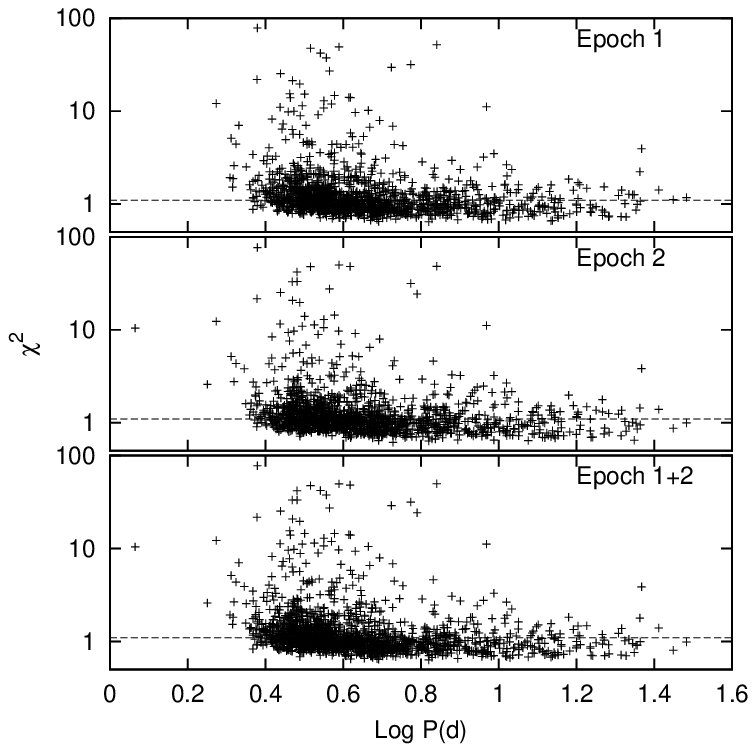}{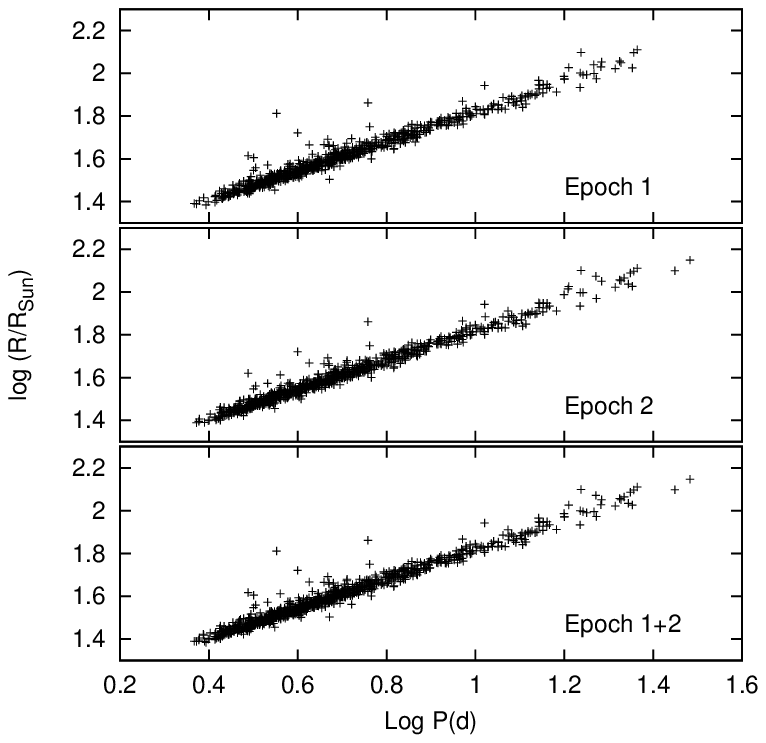}
\caption{(Left) The $\chi^2$ fit of the radius to each Cepheid in the sample for each epoch of SAGE observations plus the average of the two, with large separation Cepheids removed. The dashed line in each plot represents a value of $\chi^2 = 1.1$. (Right) The best-fit radius of each Cepheid in the sample for those Cepheids with $\chi^2<1.1$.}
	\label{f2}
	\end{center}
\end{figure*}

\section{Fitting the Radius to the Data}
To establish a reference sample, we begin by assuming no mass loss and then use Equation \ref{e1} to fit the mean radius of each Cepheid.  We fit the radius to each epoch of SAGE observations as well as the average of the fluxes for the two epochs. In Figure \ref{f2}, we show the value of $\chi^2$ and the radius for each Cepheid in each sample.  In the figure, those Cepheids with a large separation on the sky, as shown in Figure \ref{f1}, are removed; for epoch 1 this eliminates $62$ out of $1398$ Cepheids,  for the second epoch $107$ of the $1387$ Cepheids are removed, and for the average of the two epochs $73$ out of $1552$ are removed.  Only those Cepheids with a fit of $\chi^2 \le 1.1$ are shown in the plot of the predicted radius of each Cepheid.  The $\chi^2$ cut of $1.1$ for radius-only fit is taken so that the slopes of the predicted $V$ and $I$-band P-L relations agree within the uncertainties with the slopes of the observed $V$ and $I$-band P-L relations given by \cite{Ngeow2008b}.  The connection between the structures of the P-L relations and the quality of fit of the modeled data is discussed in Section 9. Fitting only the radius to the observed fluxes appears to be a reasonable model. 
\begin{figure*}[t]
\begin{center}
	\epsscale{1.1}
		\plottwo{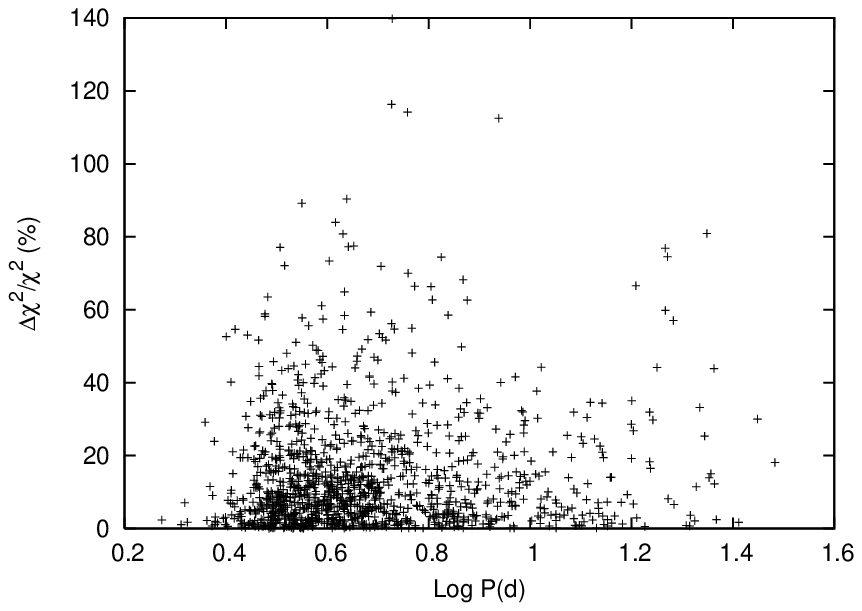}{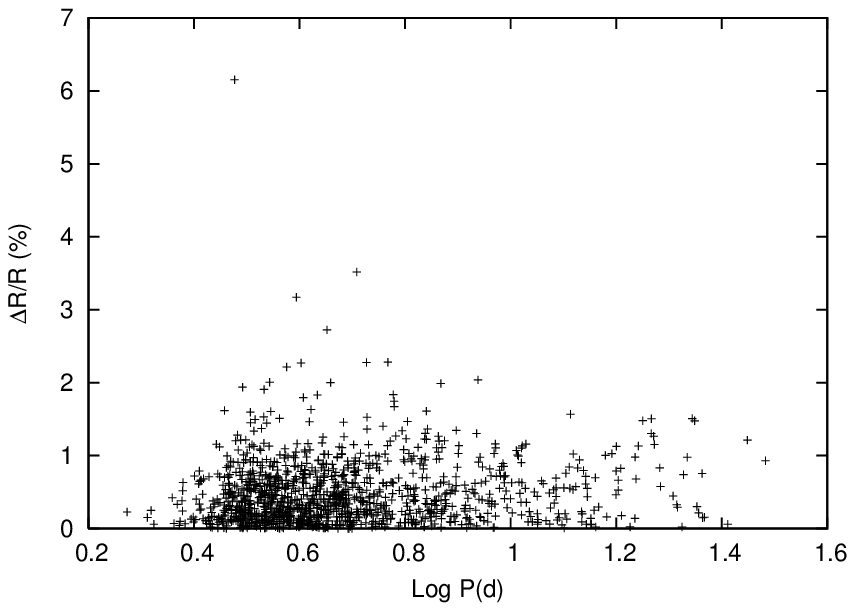}
\caption{(Left)  The percentage difference of $\chi^2$ from the radius-only model for the fits of each epoch of SAGE observation. (Right) The percentage difference of the best-fit radius of each Cepheid for each epoch of SAGE observation. }

	\label{f3}
	\end{center}
\end{figure*}

It is worth checking how much the predicted radius and $\chi^2$ value changes for each epoch. In Figure \ref{f3}, we show the relative difference of the $\chi^2$ values (Left) and the relative difference between the predicted radii (Right).  The relative difference is defined as the difference between each epoch of observation divided by the predicted quantity from the average of the two epochs.  The values of $\chi^2$ are sensitive to the differences of the infrared flux because of the increased uncertainty of the infrared fluxes due to the unknown phase of pulsation at the time of the observations and to an IR pulsation amplitude that can be up to $0.3$ magnitudes.  However, it is clear that the predicted radius is not sensitive to the differences between the infrared fluxes at each epoch of SAGE observation.  The \emph{maximum} difference of radius ranges from about $3\%$ at $35~R_\odot$ ($\log P = 0.6$) to about $2\%$ for a Cepheid with $R_* = 120R_\odot$ ($\log P = 1.4$) Cepheid. The radius depends on the total brightness and not the shape of the spectral energy distribution.  Therefore, the radius is not sensitive to changes in the infrared flux due to pulsation.

\begin{figure*}[t]
\begin{center}
	\epsscale{1.15}
		\plottwo{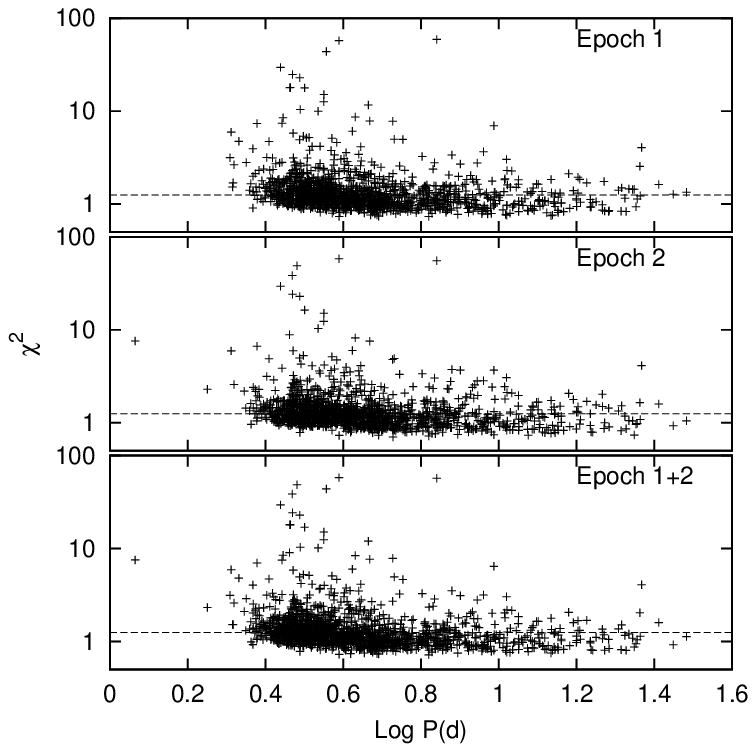}{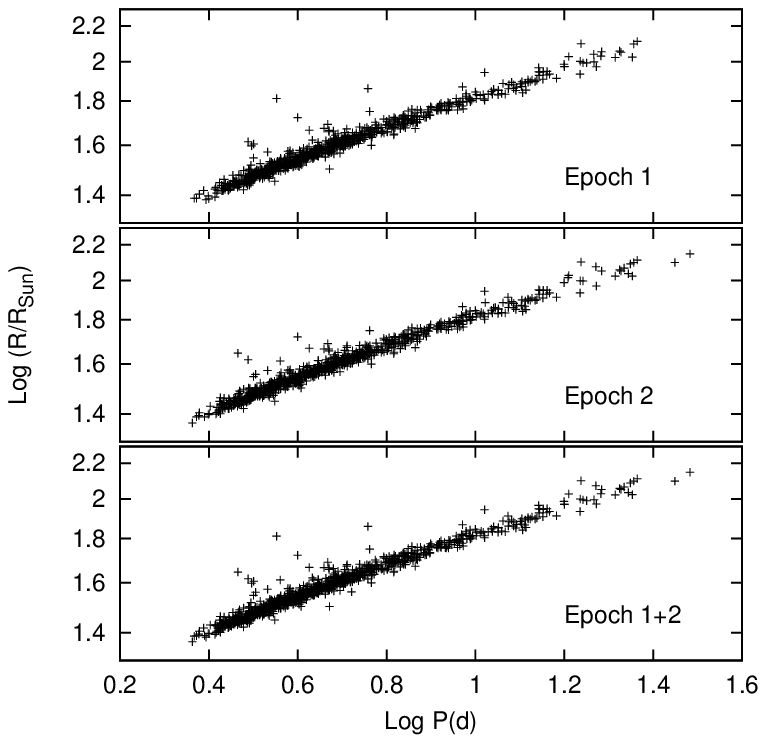}
\caption{(Left) The $\chi^2$ fit of the radius and mass-loss rate to each Cepheid in the sample for each epoch of SAGE observations plus the average of the two, with large separation Cepheids removed. The dashed line in each plot represents a value of $\chi^2 = 1.25$. (Right) The best-fit radius of each Cepheid in the sample for those Cepheids with $\chi^2<1.25$.}
	\label{f4}
	\end{center}
\end{figure*}

\begin{figure}[t]
\begin{center}
	\epsscale{1.2}

		\plotone{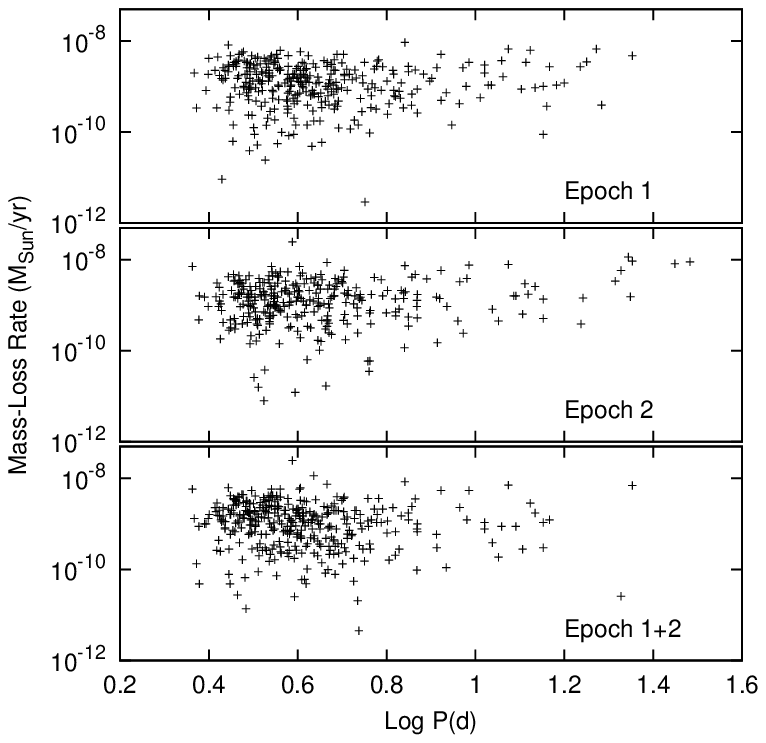}
\caption{The predicted mass-loss rates of the sample of Cepheids with $\chi^2< 1.25$ for each set of IR fluxes.}
	\label{f5}
	\end{center}
\end{figure}

\section{Fitting the Data with a Two-Parameter Model}
We now expand the model by allowing the mass-loss rates to be greater than zero, now using both the Cepheid radius and mass-loss rate in the fits.  In Figure \ref{f4}, we plot the $\chi^2$  computed for the two-parameter model, along with the predicted radius of each Cepheid with a value of $\chi^2 \le 1.25$.  There are the same number of Cepheids in this sample with $\chi^2  \le 1.25$ as there are in the fit of the radius only.  Compared to Figure \ref{f2}, the $\chi^2$-plots of Figure \ref{f4} for the two-parameter model are very similar but appear to have slightly more scatter in the fits of the long-period Cepheids. 

In Figure \ref{f5} we show the predicted mass-loss rates of the Cepheids with $\chi^2 \le 1.25$, predicted for each epoch of SAGE data.  The predicted gas mass-loss rates range from zero to $10^{-8}M_\odot/yr$, but it must be noted that these are the {\it{minimum}} mass-loss rates. The percentage of Cepheids with non-zero mass-loss rates are $45\%$ for the first SAGE epoch, $45\%$ for the second SAGE epoch, and $48\%$ for the average of the two epochs.  For many of the Cepheids with zero mass loss, a small increase of $\chi^2$ from the minimum would produce mass-loss rates of the order $10^{-10}~M_\odot/yr$.  The result appears to suggest that the unknown pulsation phase causes the infrared excess, which is discussed in the next section.  The uncertainty of the mass loss spans orders of magnitude.  The gas mass-loss rate depends on a number of assumptions, including the dust velocity, the dust-to-gas ratio, and the size and shape of dust particles. The assumed values for the dust-to-gas ratio, the dust velocity, and the dust size are all likely to be lower limits.  If the properties of any of these three are different from the assumed values then the mass-loss rates may be significantly larger.  For instance, we assumed that the dust-to-gas ratio is $1:250$, which is the maximum possible value; we noted in \cite{Neilson2008c} that the ratio may be about $1:1000$ or even smaller if the LMC metallicity is not simply scaled down from the Galactic metallicity.  This suggests that the gas mass-loss rates may be larger by the same ratio.  Also, the assumed distribution of dust grain sizes has an effect on the predicted mass-loss rates.  The dust grain sizes assumed in this work are based on the model for Galactic Cepheids \citep{Neilson2008a}, where the metallicity is larger.  In the LMC, with a smaller metallicity, it is likely the dust grain sizes have not grown as large.  If so, then the corresponding mass-loss rates need to be larger to match the observations.

\begin{figure}[t]
\begin{center}
	\epsscale{1.2}
		\plotone{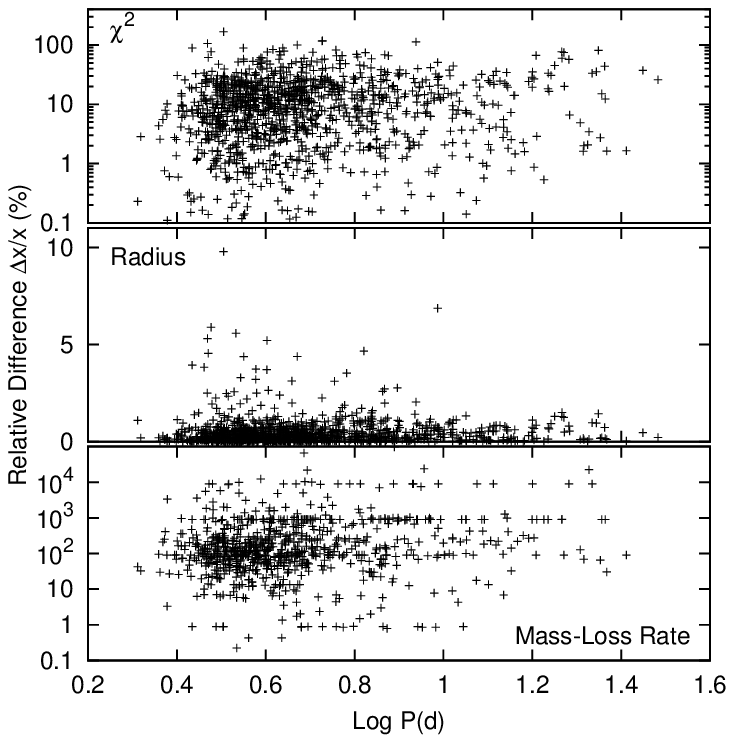}
\caption{The percentage difference between the predicted value of $\chi^2$, radius, and mass-loss rate for each Cepheid between Epoch 1 and 2 SAGE observations.}
	\label{f6}
	\end{center}
\end{figure}

Because the two SAGE epochs recorded the Cepheids at random pulsation phases, we computed the values of $\chi^2$, radius and mass-loss rates change for both epochs to assess how these variables change with phase; Figure \ref{f6} shows the percentage differences of these variables as a function of pulsation period.  The relative change of $\chi^2$ ranges from $0.1\%$ to $200\%$, suggesting that the fit is sensitive to the phase of pulsation, but this range is the same as found for the $\chi^2$ values for fitting the radius only.  The radius variation between the two epochs is typically only a few percent but some of the fits show a larger variation than in the radius-only model. The mass-loss rate shows  the greatest variation, from about $1\%$ up to $10^4\%$, approximately four orders of magnitude.  There are some Cepheids that have a mass-loss rate variation $<< 0.1\%$ and some with a mass-loss rate variation that is $>> 10^{4}\%$.  The first situation is where the predicted mass-loss rates are identical in both epochs whereas in the second case the mass-loss rate is effectively zero in one of the epochs but is significant in the other epoch.   The number of Cepheids in both cases is small.  We conclude that the unknown pulsation phase leads to a large uncertainty in the mass-loss rate because the change in infrared luminosity (in magnitudes) causes an exponential change in the mass-loss rate.

We use the F-test to compare the radius-only  model and the radius-plus-mass-loss model.  If the value of $F$ is $ > 4$ then we can state with $95\%$ confidence that the mass-loss model provides a better fit than the radius only model.  For the first SAGE epoch $F=2000$, for the second SAGE epoch $F=2800$ and for the average of the two epochs $F=3200$.  This strongly suggests that mass loss is an important phenomenon in the LMC Cepheids.

\section{Dependence of Mass Loss on the Unknown Pulsation Phase}
We have shown that slightly more than half of the Cepheids in the sample are consistent with zero mass loss, and this fraction is what one would expect if the infrared excess was entirely due to the unknown phase of the observe IR fluxes.  This expectation can be tested by varying the IR flux of a sample of $200$ Cepheids (the first $200$ in the sample, which is effectively a random period distribution).  The amplitude of the IR flux for each Cepheid is about $1/3$ the I-band amplitude from OGLE-III, and for each of the $200$ Cepheids the IRAC fluxes are allowed to vary within the amplitude.  We compute the best-fit mass-loss rate, and radius for each Cepheid with seven different phases and count how many of these phases predict significant mass-loss rates, $\dot{M} \ge 10^{-14}~M_\odot/yr$.  These seven phases create eight bins, the first being $0$ where a Cepheid has a mass-loss rate $< 10^{-14}~M_\odot/yr$ at every phase, the second bin is $1$ where a Cepheid is found to have significant mass loss for only one of the seven phases and so on to the eighth bin where a Cepheid has a mass-loss rate $\ge 10^{-14}~M_\odot/yr$ for all seven phases. 

\begin{figure*}[t] 
\begin{center}
	\epsscale{1.1}
		\plottwo{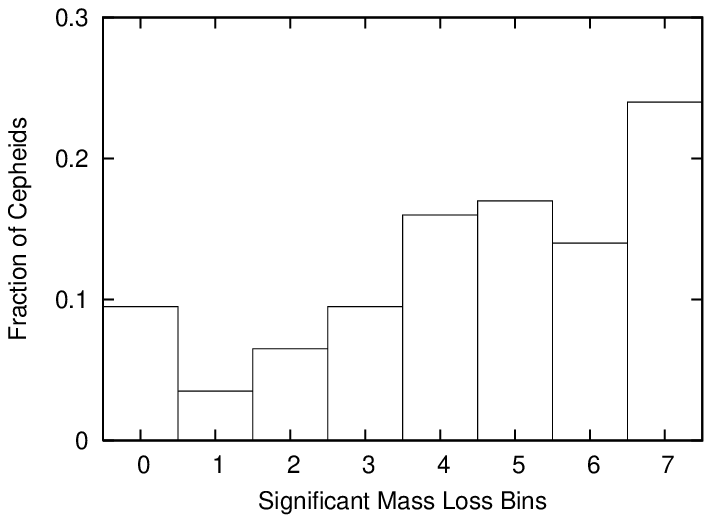}{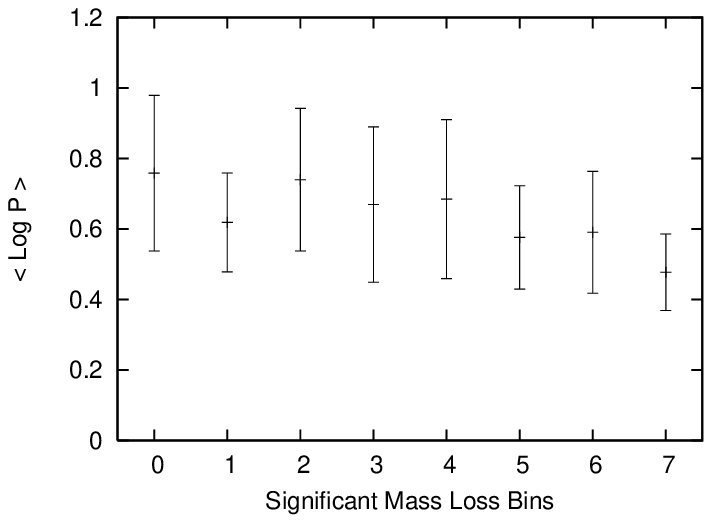}
		\caption{(Left) The number of Cepheids that have significant mass loss ($>10^{-14}~M_\odot/yr$) for a given number of phases of pulsation, where the IR flux varies.  (Right) The average period of the Cepheids in each bin.}
	\label{f7a}
	\end{center}
\end{figure*}

We plot the distribution of the Cepheids with significant mass-loss rates ($>10^{-14}~M_\odot/yr$) in Figure \ref{f7a} as well as the average period of the Cepheids in each bin.   The distribution of Cepheids appears to be highest in the last bin, with the fraction decreasing with decreasing bin until the bin representing zero mass loss at all phase with  which suggests that Cepheids predicted to have significant mass-loss rates at one phase tend to have significant mass-loss rates at all phases.  If the infrared excess was completely due to the unknown phase then one would expect that the distribution would have approximately equal peaks at each end of the distribution, i.e. for zero mass loss and mass loss at all phases.  The distribution would appear parabolic, with a maximum at zero and the minimum at bins 3, and 4, with another maximum at bin 7. However, this behavior is not seen here.  The fraction of Cepheids with mass loss at all phases is almost three times more than the fraction with zero mass loss at all phases and the distribution is skewed towards more mass loss.   This suggests that the mass-loss rates are real and not an artifact of the unknown pulsation phase, although the unknown pulsation phase is clearly an important uncertainty.

The average period of the Cepheids with mass loss at all phases raises an interesting question, where $<\log P> = 0.48$. \cite{Ngeow2008b} cut off the 5.8 $\mu$m data at $\log P = 0.536$ and cut off the 8.0 $\mu$m data at $\log P = 0.702$. The average period of the Cepheids in all bins except the $0$ bin is less than the $8~\mu m$ cut-off period while bin 7 has an average period less than the $5.8~\mu m$ cut-off period.  This is not surprising; \cite{Ngeow2008b} applied the cut-off periods to remove Cepheids near the limiting magnitude of the observations at these wavelengths.  This result does suggest we test the propensity for mass loss with $5.8$ and $8.0~\mu m$ data removed from the calculations.

\begin{figure*}[t] 
\begin{center}
	\epsscale{1.15}
		\plottwo{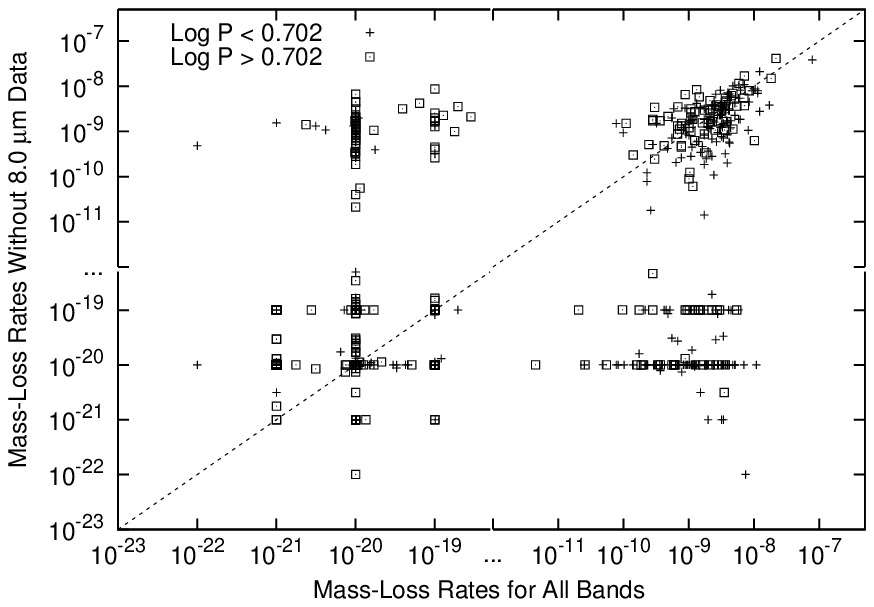}{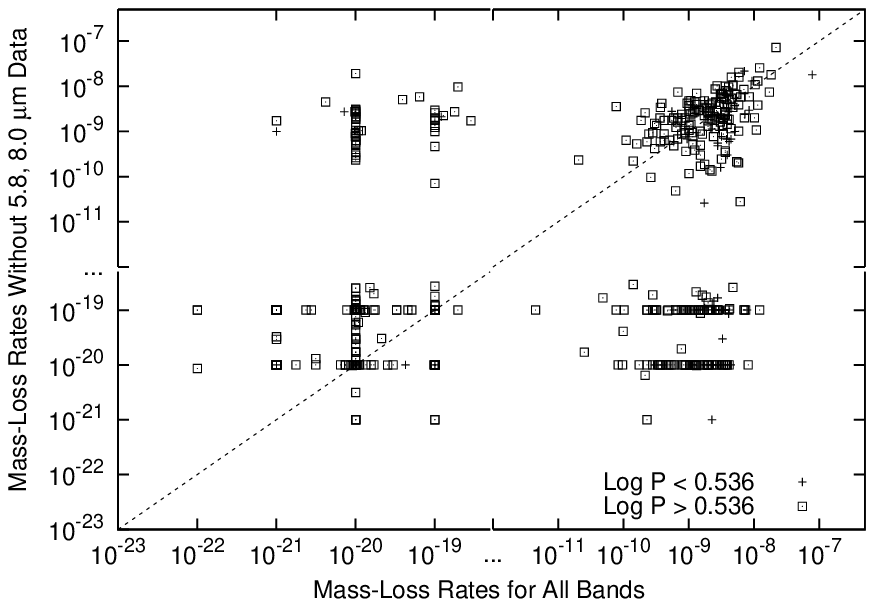}
		\caption{(Left) The comparison of mass-loss rates with the $8.0~\mu m$ data removed and mass-loss rates using all wavebands. (Right) Comparison of mass-loss rates with $5.8$ and $8.0~\mu m$ data removed and mass-loss rates using all wavebands.  The dotted line in both plots denotes a slope of unity. In both plots there is an abrupt jump from $10^{-19}$ to $10^{-11}~M_\odot/yr$.  Predicted mass-loss rates $\le 10^{-19}~M_\odot/yr$ refer to predictions of negligible mass loss. The abrupt jump is inserted because there are no Cepheids with predicted mass-loss rates in that range.}
	\label{f7}
	\end{center}
\end{figure*}

Our first test recalculates the best fit radius and mass-loss rates after removing the 8.0 $\mu$m data from the sample, and compares this to the mass-loss rates calculated using fluxes from all wavelengths; the comparison is shown in Figure \ref{f7}. The data are plotted for two groups: those Cepheids with $\log P < 0.702$ and those with longer periods. The mass-loss rates of both populations have similar scatter when the 8.0 $\mu$m fluxes are excluded. From this we conclude that photometric errors in the 8.0 $\mu$m  fluxes at the detection limit do not significantly affect the  mass-loss model.  Short-period Cepheids with 8.0 $\mu$m brightness of about $12$ to $11$ magnitudes (see Figure 3 of \cite{Ngeow2008b}) have mass-loss rates that vary by about a factor of only two when that flux is used to constrain the model.  It is these Cepheids that appear to have the most significant IR excesses, yet the mass-loss rates are not affected by photometric errors.  It is possible that the short-period Cepheids with $8.0$ $\mu m$ brightnesses of about $14$ to $13$ mag. are affected by photometric errors at the limiting magnitude but there is similar scatter for longer period Cepheids.  It is not possible to quantify how much this effect contributes, if any, but we can state that the contribution does not affect the results significantly.  

Next we recalculate the best fit radius and mass-loss rates with both the 5.8 and the 8.0 $\mu$m data removed, and we have grouped the data into two bins, those with periods $\log P< 0.536$ and those greater.  Again, we compare to the radii and mass-loss rates calculated using all the wavelengths; the results are shown in the right panel of Figure \ref{f7}.  The scatter in the mass-loss rates is similar, and the exclusion of the 5.8 and 8.0 $\mu$m data has not made a substantial reduction in the number of Cepheids predicted to have significant mass loss rates. The apparent rows at low mass-loss rates in Figure \ref{f7} are due to the stopping the calculations when it is clear the the best-fit mass-loss rate in the analysis is less  than $10^{-15} M_{\odot}/yr$, which is equivalent to zero mass loss  in the analysis. The effective temperature is determined by the period and $(V-I)_0$, and the radius is most dependent on the optical and near-infrared fluxes, which implies that the stellar luminosity is well determined. The IRAC fluxes provide the strongest constraint on the mass-loss rates, indicating that the 3.6, 4.5, and 5.8 $\mu$m fluxes help offset this effect at 8.0 $\mu$m.  When we compute the IR P-L relations, we will reconsider the influence of the 8.0 $\mu$m flux limit.
   
\section{Predictions of the Period-Radius Relation}

Because one of the free parameters of our model is the stellar radius, the fitting procedure determines the radius of each Cepheid in the sample, enabling us to create a Period-Radius (P-R) relation for LMC Cepheids.  We derive two P-R relations, one using  the simpler model in which the radius as the only free parameter,  and a second P-R relation in which both radius and mass loss are included.  Table \ref{t1} gives the slopes and zero-points of the P-R relations for each of the two models and for each epoch of the SAGE  data.  The P-R relation is expressed in the form  $\log R(R_{\odot})= a \log P(d) + b$ determined using a $3\sigma$ 
iterative clipping routine, similar to that used by \cite{Ngeow2008b}. For the fit using just the radius, the slopes and zero-points of the P-R relations for various combinations of SAGE epochs all agree within the uncertainty.  Similarly, the slopes and zero-points of each P-R relation using the mass-loss model for the three SAGE epochs agree. A comparison of the predicted P-R relations from the radius-only model to the P-R relations found using the mass-loss model shows that the slopes and zero-points  agree within the uncertainty, albeit the zero-point of the radius-only model is marginally larger but well within the uncertainty due to distance to the LMC. Note that the unknown phase of the infrared observations does not affect the fit of the P-R relation because the affect of phase is random for each Cepheid and cancels out in the fit.
\begin{table}[t]
\caption{Best Fit Parameters for Predicted Period-Radius Relations}
\begin{center}
\begin{tabular}{llccccc}
\hline
 Epoch & Slope & Zero Point & Dispersion & N & F\\
\hline
\multicolumn{5}{c}{Radius Model}\\
\hline
 1 &$0.690 \pm 0.003$ & $1.129 \pm 0.002$ & $0.017$ & $692$&$3.57$\\
  2&$0.688\pm 0.003$&$1.131\pm 0.002$&$0.017$& $700$&$0.66$\\
 1+2 & $0.689\pm 0.003$&$1.130\pm 0.002$ & $0.017$& $817$ &$0.88$\\
 \hline
\multicolumn{5}{c}{Radius $+$ Mass-Loss Model}\\
\hline
 1 &$0.692 \pm 0.003$ & $1.127 \pm 0.002$ & $0.017$ & $ 701$ &$3.16$\\
 2&$0.688\pm 0.003$&$1.130\pm 0.002$&$0.017$ & $710$&$0.63$\\
 1+2 & $0.690\pm 0.003$&$1.128\pm 0.002$ & $0.017$ & $822$&$0.74$\\
\hline
\end{tabular}
\end{center}
\label{t1}
\end{table}

Table \ref{t1a} gives the P-R relations from several other investigations.
 \cite{Gieren1999} used the Infrared Surface Brightness technique to determine radii of LMC and SMC Cepheids,  \cite{Groenewegen2007}  derived a P-R relation from five galactic Cepheids with measured distances and angular diameters, while  \cite{Kervella2004b} used the surface brightness method as well. \cite{Ruoppo2004} used the CORS method (a modified version of the Baade-Wesselink method derived by \cite{Caccin1981}) and \cite{Gieren1998} used the IRSB technique  to find $\log R = 0.750(\pm0.024)\log P + 1.075(\pm 0.007)$ that is identical to the results of \cite{Laney1995} using the surface brightness method.   \citep{Bono1998}derived P-R  relations from their theoretical calculations for two different mass-luminosity ($M-L$) relations; the shallower $M-L$ relation the larger slope for the P-R relation.  Our results agree well
with the relations found by \cite{Gieren1999},  \cite{Ruoppo2004}, 
and \cite{Groenewegen2007}, but not with the relations with slopes $> 0.750$. Our results using both the one- and two-parameter fits predict Cepheii radii that are consistent with results from 
observations and with the P-R relations.

\begin{table*}[t]
\caption{Period-Radius Relations From the Literature}
\begin{center}
\begin{tabular}{lcc}
\hline
Reference & Slope & Zero Point \\
\hline
\cite{Gieren1998} & $0.750\pm 0.024$ & $1.075\pm 0.007$\\
\cite{Bono1998}, Canonical & $0.666\pm 0.007$&$1.192\pm0.009$\\
\cite{Bono1998}, Non-Canonical & $0.653\pm 0.006$&$1.183\pm0.009$\\
\cite{Gieren1999} & $0.680 \pm 0.017$ & $1.146\pm 0.025$ \\
\cite{Kervella2004b} & $0.767\pm 0.009$ & $1.091\pm 0.011$\\
\cite{Ruoppo2004} & $0.69\pm 0.09$& $1.22\pm0.08$ \\
\cite{Groenewegen2007} & $0.686\pm 0.036$&$1.134\pm 0.034$ \\
\hline
\end{tabular}
\end{center}
\label{t1a}
\end{table*}

As an experiment, we tested our sample for the possibility of non-linear P-R relations, define as two linear relations with a period break at P = 10 days, similar to non-linear P-L relations defined in  \cite{Kanbur2004}. There is no physical reason to expect that the relations should be non-linear, but it is a useful check of the fitting models. We do not list the non-linear P-R relations, but have tested the possibility using the F-test again. Because there are more than 1000 Cepheids in each sample, a value of $F \geq 3$ indicates the relations are consistent with being non-linear. The results from the SAGE epoch 2 and average data are consistent with a linear P-R relation but the P-R relations found using the mass-loss model and radius-only model from the first SAGE epoch data are consistent with being non-linear, with $F = 3.16$ for the mass-loss model and $F = 3.57$ for the radius-only model.  This anomaly is interesting but not sufficient evidence for non-linearity of the P-R relation.

\section{The Effect of Mass Loss on the Infrared Surface Brightness Method}
 A number of the Period-Radius relations cited in the previous section use the Infrared Surface Brightness method to determine the angular diameter of the Cepheids.  However, if mass loss is important and causes infrared excess, the the observed infrared fluxes will be brighter than the star alone.  In this case, mass loss will distort the angular diameters predicted by near-infrared surface brightness relations from \cite{Fouque1997},  who found
 \begin{eqnarray}
 \log \theta & = & 0.5474 - 0.2V_0 + 0.262(V - K)_0, \label{e4}\\
 \log \theta & = & 0.5474 - 0.2K_0 + 0.220(J - K)_0.  \label{e5}
 \end{eqnarray}
 
The IRSB technique is calibrated with individual Cepheids, and if  these Cepheids have mass loss causing an infrared excess, this will distort the predictions of the methods.  Specifically, four Cepheids used in the calibration, V Cen, U Sgr, RZ Vel, and U Car, are predicted to have mass-loss rates ranging from $10^{-10}$ to $10^{-8} M_{\odot}/yr$ \citep{Neilson2008a}. These mass-loss rates could increase the uncertainty of the coefficient multiplying the color term, causing a greater scatter in the fitting of the calibration sources, and possibly causing the coefficient to be overestimated. 

The infrared excess from mass loss will also affect the value of the resultant angular diameter in two ways. First, it will cause the Cepheid to appear brighter in the $J$-band, and even more so in the $K$-band. Second, the brighter $J$- and $K$- magnitudes will increase the $(V-K)_0$ and $(J-K)_0$ colors, further increasing the angular diameters. Therefore, mass loss tends to predict larger angular diameters of Cepheids and hence larger radii when the Baade-Wesselink method is employed for the $(V-K)_0$ relation.  It is unclear how the $(J-K)_0$ relations is affected because the infrared excess makes the Cepheid brighter in $J$-band which acts to counter the increase of the color.

We begin by assessing how mass loss affects the uncertainty of the coefficient of the color terms in the relations.  To fit the stellar radius to the observations, we have assumed a distance to each Cepheids, which is equivalent to fitting each Cepheid's angular diameter.  The predicted radii and angular diameters derived from the mass-loss model differ by less than a few percent from the predicted radii and angular diameters from fitting only the radius to the observations. This difference between the radii has a negligible effect on the comparison of the angular diameters from the IRSB technique.

\begin{figure*}[t]
\begin{center}
	\epsscale{1.15}
		\plottwo{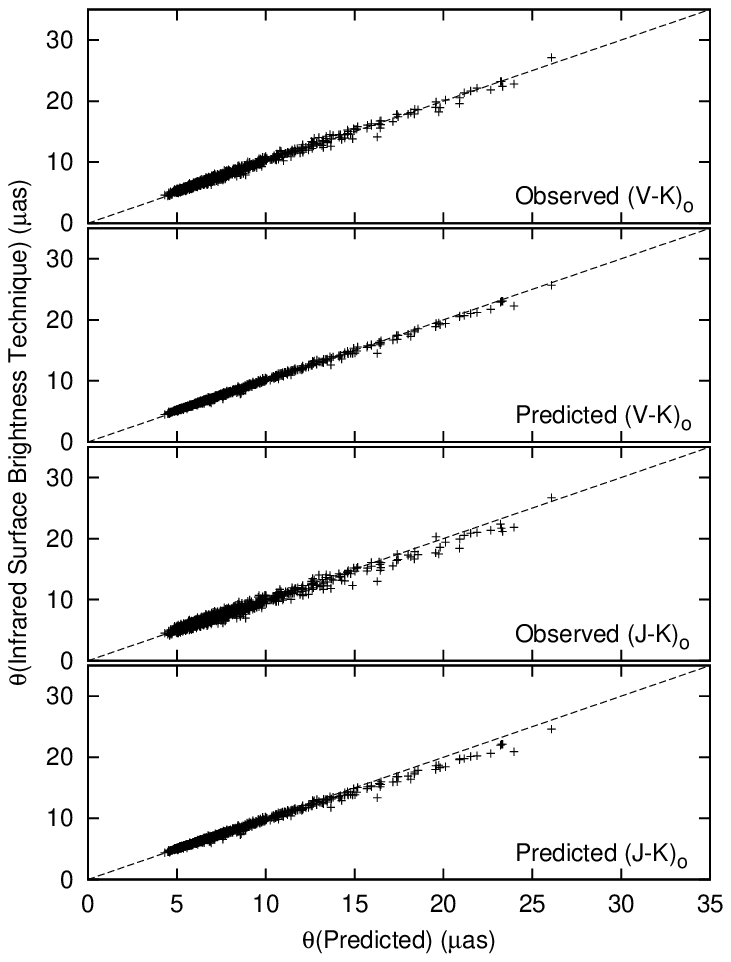}{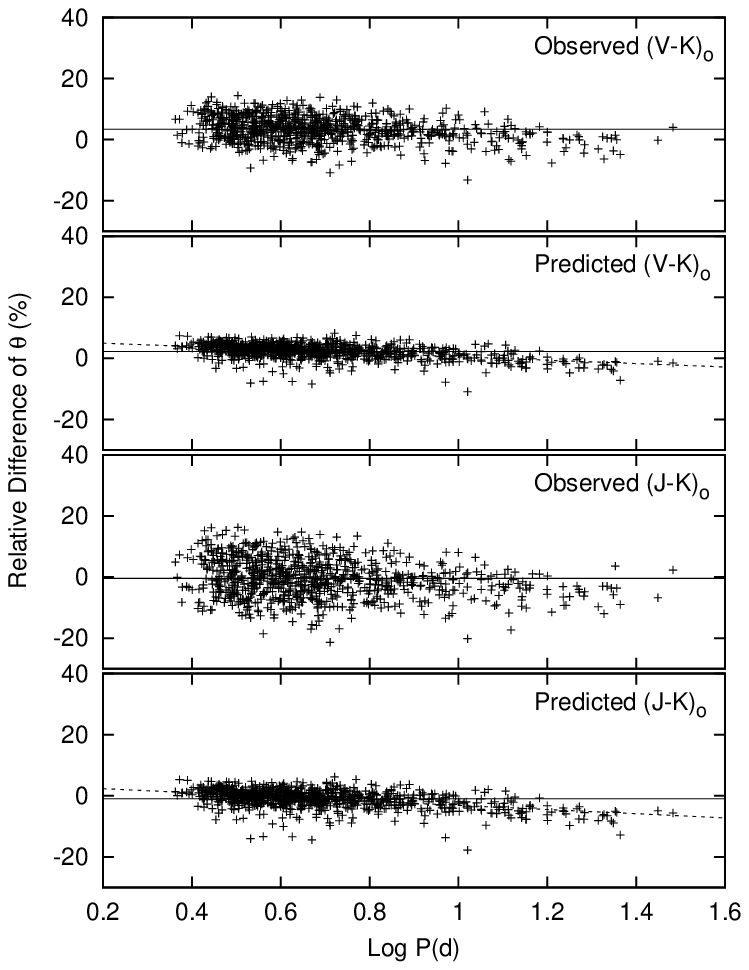}
\caption{(Left) Comparison of the angular diameters of the Cepheids using the best-fit radius and a distance modulus of $18.5$ and the angular diameters found by using the IR surface brightness technique of \cite{Fouque1997}, Equations \ref{e4} and \ref{e5}, using the predicted stellar fluxes and the observed fluxes from the average of the two SAGE epochs with a fit $\chi^2 \le 1.25$. The dashed lines represent where the predicted angular diameter equals the angular diameter from the IR surface brightness method.  (Right) The fractional difference between the IR surface brightness technique and the best-fit radius as a function of pulsation period.  The solid horizontal lines represent the average fractional difference of the sample and the dotted lines (in the plots using the predicted stellar fluxes) are best-fit linear relations.}
	\label{f9}
	\end{center}
\end{figure*}

The mass loss also affects the predicted angular diameter via the $J$- and $K$-band observations.  This influence is included when the angular diameters of the Cepheids are computed from the observed fluxes from the \cite{Ngeow2008b} data set, and from the predicted stellar fluxes found by fitting the mass-loss model, using Equations \ref{e4} and \ref{e5}. Figure \ref{f9} compares the angular diameters using the IRSB technique using observed fluxes and the angular diameters derived using the predicted fluxes with the best-fit angular diameter from the mass-loss model, which is the best-fit radius divided by the assumed distance to the LMC.  The angular diameters from the observed fluxes exhibit significantly more scatter than the angular diameters from the predicted fluxes. There is reasonable agreement between the angular diameters from the predicted fluxes from Equation \ref{e4} and \ref{e5}, although the comparison from Equation \ref{e5} has a larger range of differences. 

In the right panel of Figure \ref{f9}, we show the relative difference between the angular diameters predicted from the observed and predicted $V$-, $J$-, and $K$-band fluxes and the best-fit angular diameters as a function of pulsation period.  The solid lines represent the mean relative difference and the dotted lines are the best-fit linear relation.  The average difference between the angular diameter found using the observed $(V-K)_0$ and the best-fit angular diameter is $3.35\%$, while the predicted stellar $(V-K)_0$ gives a difference of $2.26\%$.  Doing the comparison of the best-fit angular diameter found with the observed $(J-K)_0$, the relative difference is $-0.42\%$, while using the predicted $(J-K)_0$ gives a relative difference of  $-0.99\%$.  The lower average uncertainty for the $(J-K)_0$ comparison is likely due to the mass loss having a similar effect on both the $J$- and $K$-band magnitudes, partially canceling out the effect of the color term.  The $K$-band magnitudes will also be smaller when infrared excess is not removed, thereby predicting a smaller angular diameter, which explains why the average difference is negative. The relative difference between the angular diameters is a measure of the uncertainty of  the IRSB calibration technique, which suggests that the IRSB technique overestimates the radius of a Cepheid by a few percent because of the correlation present in the coefficient multiplying the color.  The differences between the angular diameters derived from the observed fluxes and the best-fit angular diameters measure the uncertainty of both the calibration and the uncertainty due to infrared excess.  The average uncertainty due to infrared excess is about $3$ to $4.5\%$ on average, although in rare cases the uncertainty may be up to $10$ to $20\%$.  

Figure \ref{f9} displays a slight period dependence.  One possible cause of this dependence is if the stellar color is not solely related to the effective temperature.  There is also a deviation from linearity for small periods.  This deviation could be due to the interaction of the photosphere with the hydrogen ionization front \citep{Simon1993}, which would make the surface brightness 
relation (Equations \ref{e4} and \ref{e5}) nonlinear functions of color.  Because the ionization front is closer to the photosphere for hotter Cepheids, this interaction has a greater effect for shorter-period Cepheids, which alters the effective temperature from that of a non-pulsation star of the same mass, luminosity and radius.  For longer-period Cepheids the interaction is diminished because the ionization front is deeper; the effective temperature and color are more similar to a non-pulsating star. Because the calibration Cepheids used for the IRSB method may be preferentially shorter period stars, this could lead to systematic differences when applied to longer-period Cepheids.  This period dependence is similar to the dependence of the distance modulus of the LMC found by \cite{Gieren2005} using a constant value of the projection factor.  In order to obtain a constant distance modulus, \cite{Gieren2005} derived a period-projection factor relation that is different from that determined by \cite{Nardetto2007} from spectroscopy.   It is possible that mass loss and IR excess contribute to the difference.

\section{Predictions of the Linear Period-Luminosity Relations}

We have shown that mass loss leads to infrared excess that changes the $J$- and $K$-band fluxes of Cepheids, which, in turn, affect the angular diameters determined from the IRSB method.  The infrared excess may also affect the infrared P-L relations.

In investigating the infrared P-L relations, we continue to work with the $VIJHK$ data from two SAGE epochs as well as the IRAC data. Because the infrared excess due to mass loss will not contribute flux to the $V$- and $I$-bands, the slopes of the predicted $V$- and $I$-band P-L relations should be the same as the observed relations.  However, the structure of the P-L relations will be influenced by the value of $\chi^2$ chosen for the threshold; here we take $\chi^2 < 1.25$ as our limit.  The zero-points may vary because the $V$- and $I$-band fluxes depart from perfect blackbodies, but the uncertainty is about 0.1 mag, shown in our previous article. The predicted linear relations are computed using a $3 \sigma$-clipping algorithm for the SAGE data from each epoch separately as well as the average of the two epochs; these are shown in Table 3. There is not a large difference between the predicted slopes and zero-points found by fitting the radius only and those determined from the mass-loss model, but these differences may be attributed to the differences between the best-fit radius from each model as seen in the P-R relations shown in Table \ref{t1}. 
\begin{table*}[t]
\caption{Best Fit Parameters for Predicted Linear Period-Luminosity Relations}
\begin{center}
\begin{tabular}{cccccccccc}
\hline\hline
\multicolumn{5}{c}{Radius Model}&\multicolumn{5}{c}{Mass-Loss Model}\\
\hline
\hline
Band & Slope & Zero Point & Dispersion & N &Band & Slope & Zero Point & Dispersion & N \\
\hline
\multicolumn{10}{c}{Epoch 1} \\
\hline
V & $-2.745\pm 0.039$&$ 17.235\pm 0.028$&$0.204$&$697$&V&$-2.765\pm 0.039$&$17.252\pm 0.028$&$ 0.204$&$707$\\
I& $-2.932\pm 0.028$&$16.420 \pm0.020$&$ 0.148$&$698$&I&$-2.947\pm 0.028$&$ 16.433\pm 0.020$&$ 0.148$&$ 707$\\
J& $-3.081\pm 0.021$&$16.298\pm 0.015$&$ 0.112$&$ 701$&J&$-3.099\pm 0.021$&$16.313\pm 0.015$&$0.109$&$ 706$\\
H& $-3.135\pm 0.019$&$16.072\pm 0.014$&$ 0.100$&$702$&H&$-3.147\pm 0.019$&$16.084\pm 0.013$&$ 0.099$&$ 709$\\
K& $-3.181\pm  0.017$&$15.993\pm 0.013$&$ 0.091$&$703$&K& $-3.193\pm 0.017$&$ 16.005\pm 0.012$&$0.089$&$710$\\
$3.6$& $-3.240 \pm 0.015$&$15.967 \pm 0.011$&$ 0.080$&$698$&$3.6$&$ -3.250\pm 0.015$&$15.978\pm 0.011$&$ 0.079$&$ 705$\\
$4.5$&$-3.253\pm 0.015$&$15.894\pm 0.011$&$0.079$&$698$&$4.5$&$-3.266\pm 0.015$&$ 15.906\pm 0.011$&$ 0.078$&$ 706$\\
$5.8$&$-3.265\pm 0.015$&$15.896\pm 0.011$&$0.077$&$698$&$5.8$&$-3.279\pm 0.016$&$ 15.910\pm 0.011$&$ 0.077$&$ 707$\\
$8.0$&$-3.275\pm 0.015$&$ 15.899\pm 0.011$&$0.076$&$697$&$8.0$&$-3.288\pm 0.014$&$15.913\pm 0.010$&$ 0.076$&$706$\\
 \hline
\multicolumn{10}{c}{Epoch 2} \\
\hline
V&$-2.778 \pm 0.037$&$ 17.265 \pm 0.027$&$ 0.200$&$704$&V&$-2.776\pm 0.036$&$17.267\pm 0.026$&$0.197$&$713$\\
I&$-2.958\pm 0.027$&$ 16.440\pm 0.019$&$ 0.144$&$ 704 $&I&$-2.964\pm 0.026$&$16.449\pm 0.019$&$0.144$&$716$\\
J&$-3.102\pm 0.020$&$ 16.314\pm 0.014$&$0.109$&$708$&J&$-3.105\pm 0.020$&$16.320\pm 0.014$&$0.108$&$718$\\
H&$-3.148\pm 0.018$&$16.083\pm0.013$&$0.099$&$712$&H&$-3.151\pm 0.018$&$16.088\pm 0.013$&$ 0.985$&$722$\\
K&$-3.198\pm0.016$&$16.005\pm 0.012$&$0.091$&$714$&K&$-3.200\pm 0.016$&$ 16.009\pm 0.012$&$ 0.090$&$724$\\
$3.6$&$-3.251\pm 0.015$&$ 15.975\pm 0.011$&$ 0.080$&$708$&$3.6$&$-3.251\pm 0.015$&$ 15.977\pm 0.010$&$0.079$&$ 715$\\
$4.5$&$-3.260\pm 0.014$&$15.899\pm 0.011$&$ 0.080$&$706$&$4.5$&$ -3.261\pm 0.014$&$ 15.902\pm 0.010$&$ 0.078$&$ 715$\\
$5.8$&$-3.271 \pm 0.014$&$15.900\pm0.010$&$0.077$&$706$&$5.8$&$-3.272\pm 0.014$&$15.904\pm 0.010$&$ 0.077$&$715$\\
$8.0$&$-3.282\pm 0.014$&$15.904\pm 0.010$&$0.076$&$706$&$8.0$&$-3.283\pm 0.014$&$15.907\pm 0.010$&$ 0.076$&$715$\\
\hline
\multicolumn{10}{c}{Average of Two Epochs} \\
\hline
V&$ -2.720\pm 0.034$&$17.218\pm 0.024$&$0.198$&$821$&V&$-2.755\pm 0.034$&$ 17.225\pm 0.024$&$ 0.199$&$ 826$\\
I&$-2.935\pm 0.025$&$16.422\pm 0.018$&$0.143$&$821$&I&$-2.952\pm 0.025$&$16.441\pm 0.018$&$0.145$&$ 828$\\
J&$-3.075\pm 0.018$&$16.294\pm 0.013$&$0.108$&$825$&J&$-3.099\pm 0.019$&$ 16.318\pm 0.013$&$ 0.111$&$835$\\
H&$ -3.129\pm 0.017$&$16.069\pm 0.012$&$0.099$&$831$&H&$-3.144\pm 0.017$&$16.084\pm 0.012$&$ 0.098$&$835$\\
K&$-3.175\pm 0.015$&$ 15.990\pm 0.011$&$0.089$&$832$&K&$-3.189\pm 0.015$&$ 16.003\pm 0.011$&$0.089$&$ 836$\\
$3.6$&$-3.233\pm 0.014$&$15.963\pm 0.010$&$ 0.079$&$826$&$3.6$&$-3.245\pm 0.013$&$ 15.976\pm 0.010$&$ 0.078$&$828$\\
$4.5$&$-3.242\pm 0.013$&$15.887\pm 0.010$&$ 0.077$&$825$&$4.5$&$-3.253\pm 0.013$&$15.898\pm 0.009$&$ 0.076$&$ 826$\\
$5.8$&$-3.256\pm0.013$&$15.891\pm 0.009$&$0.076$&$ 824$&$5.8$&$-3.269\pm 0.013$&$15.904\pm 0.009$&$ 0.076$&$ 829$\\
$8.0$&$ -3.265\pm 0.013$&$ 15.894\pm 0.009$&$0.076$&$825$&$8.0$&$-3.280\pm 0.013$&$ 15.908\pm 0.009$&$0.075$&$829$\\
\hline \\ \\ 
\end{tabular}
\end{center}
\label{t2}
\end{table*}

To test if the signature of the infrared excess in the P-L relation is impacted by the SAGE detection limit, we have removed the 8.0 $\mu$m data and recalculated the relations.  Consistent with what we have found in previous sections, the predicted radii and mass-loss rates are not significantly changed, and the slopes of the IRAC P-L relations agree with the results in Table \ref{t2} to within the uncertainty.  This suggests that the observed brighter fluxes for Cepheids with periods $\log P < 0.7$ are real and not solely caused by being near the detection limit of the survey.

About one-half of the Cepheids in the sample are predicted to have mass-loss rates $(>10^{-14}~M_\odot/yr)$. To test the impact of these Cepheids on the IRAC P-L relation, they have been removed from the sample and the P-L relations have been recomputed for the average of the two SAGE epochs; the results are listed in Table \ref{t4}, where no period cut has been applied to the data, as in \cite{Ngeow2008b}. The IRAC P-L relations are similar to the results of  \cite{Ngeow2008a} and \cite{Ngeow2008b}, with the slopes having the same wavelength dependence, becoming shallower with increasing wavelength.   The similarity is due to the influence of the uncertainty of the IR fluxes caused by the unknown pulsation phase in the fitting process in the mass-loss model, for which a small change of IR flux might produce mass-loss rates orders of magnitude different.   The data are likely still being influenced by IR excess.
\begin{figure*}[t]
\begin{center}
	\epsscale{1.15}
		\plottwo{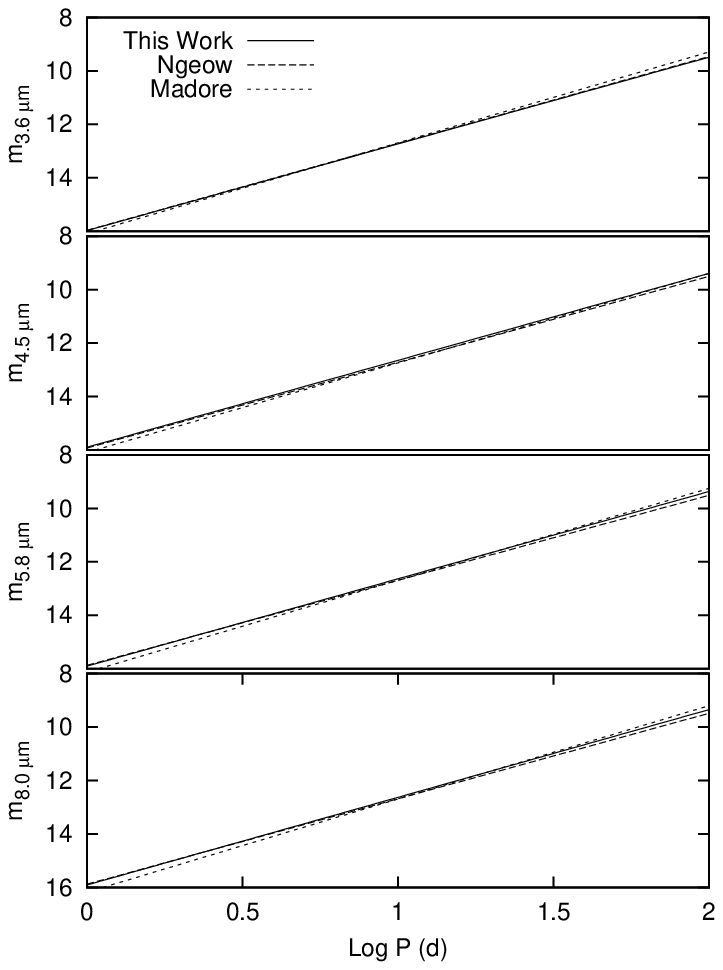}{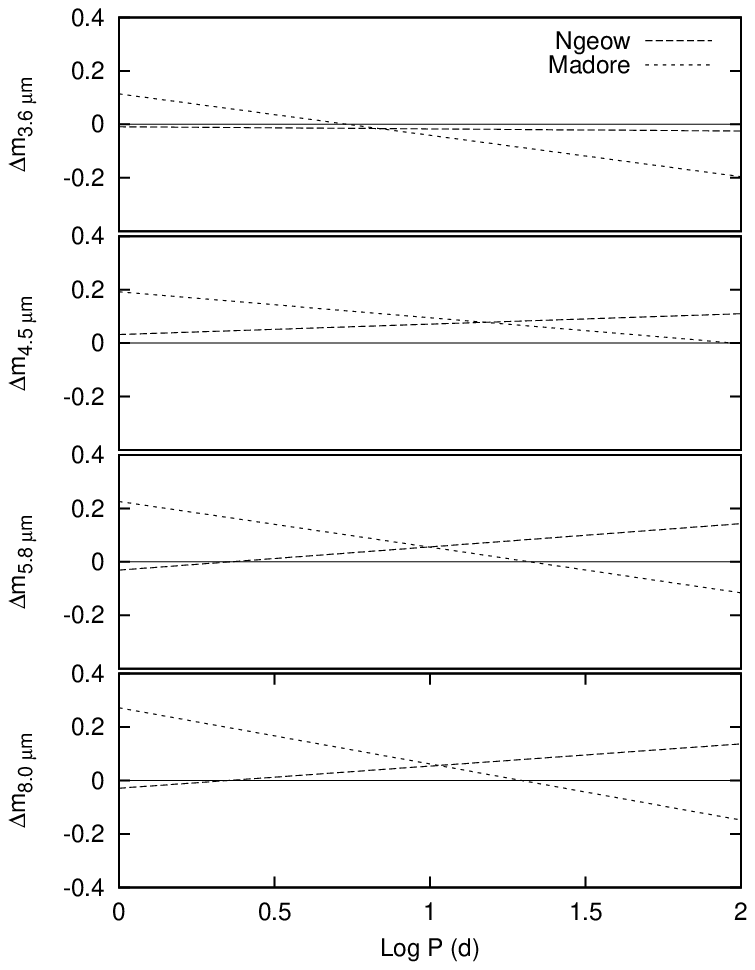}
\caption{(Left) Comparison of the IR Period-Luminosity relations from here, \cite{Ngeow2008b} and \cite{Madore2008}. (Right) The difference between the IR P-L relations of \cite{Ngeow2008b} and \cite{Madore2008} and the IR P-L relations in this work.  The horizontal line represents zero difference.}
	\label{f10a}
	\end{center}
\end{figure*}
\begin{table*}[t]
\caption{Best Fit Parameters for Observed IRAC Linear Period-Luminosity Relations}
\begin{center}
\begin{tabular}{ccccc}
\hline
Band & Slope & Zero Point & Dispersion & N \\
\hline
\multicolumn{5}{c}{Observed IR Fluxes With Mass-Losing Cepheids Removed} \\
\hline
$3.6$& $-3.224 \pm 0.015$&$15.984 \pm 0.011$&$ 0.089$&$761$\\
$4.5$&$-3.190\pm 0.015$&$15.945\pm 0.011$&$0.089$&$763$\\
$5.8$&$-3.203\pm 0.020$&$15.938\pm 0.015$&$0.120$&$763$\\
$8.0$&$-3.080\pm 0.033$&$ 15.818\pm 0.028$&$0.134$&$429$\\
 \hline
\multicolumn{5}{c}{Observed IR Fluxes With $\log P > 1.05$ Mass-Losing Cepheids Removed} \\
\hline
$3.6$& $-3.239 \pm 0.160$&$16033 \pm 0.191$&$ 0.135$&$62$\\
$4.5$&$-3.068\pm 0.172$&$15.849\pm 0.206$&$0.146$&$63$\\
$5.8$&$-3.168\pm 0.166$&$15.931\pm 0.198$&$0.140$&$63$\\
$8.0$&$-3.256\pm 0.175$&$ 16.016\pm 0.175$&$0.148$&$63$\\
 \hline
\multicolumn{5}{c}{Observed IR Fluxes} \\
\hline
$3.6$&$-3.266\pm 0.010$&$ 15.991\pm 0.007$&$ 0.100$&$1700$\\
$4.5$&$-3.210\pm 0.010$&$15.938\pm 0.007$&$ 0.102$&$1715$\\
$5.8$&$-3.072 \pm 0.017$&$15.775\pm0.012$&$0.158$&$1521$\\
$8.0$&$-2.854\pm 0.031$&$15.500\pm 0.026$&$0.221$&$815$\\
\hline
\multicolumn{5}{c}{Observed IR Fluxes for $\log P >1.05$} \\
\hline
$3.6$&$-3.347\pm 0.068$&$16.112\pm 0.087$&$ 0.150$&$118$\\
$4.5$&$-3.320\pm 0.067$&$16.093\pm 0.087$&$ 0.150$&$118$\\
$5.8$&$-3.395\pm0.069$&$16.42\pm 0.089$&$0.153$&$ 119$\\
$8.0$&$ -3.517\pm 0.074$&$ 16.264\pm 0.096$&$0.165$&$119$\\
\hline \\ \\ 
\end{tabular}
\end{center}
\label{t4}
\end{table*}

\cite{Freedman2008} and \cite{Madore2008} presented IR P-L relations for LMC Cepheids using a selected data set that is very different than the IR P-L relations found by \cite{Ngeow2008a}, \cite{Ngeow2008b} and predicted by \cite{Neilson2008c}.   \cite{Freedman2008} used SAGE epoch 1 observations and IRAC data for a sample of Cepheids from \cite{Persson2004} to derive P-L relations with slopes ranging from about -3.30 to -3.44. \cite{Madore2008} used the same Cepheid sample and the average of SAGE epoch 1 and 2 observations to derive very similar slopes ranging from $-3.35$ to $-3.49$. As shown in Figure \ref{f10a}, the slopes from \cite{Madore2008} are significantly steeper than those found here or by \cite{Ngeow2008a} or \cite{Ngeow2008b}.  There are two reasons why the observed P-L relations differ.  First, \cite{Ngeow2008a} and \cite{Ngeow2008b} used an iterative fitting routine designed to minimize the effect of Cepheids with fluxes that are very different from the majority of the sample, possibly due to effects such as blending or false identification. The method retains all the Cepheids but changes their weighting in fitting the P-L relation, thus affecting the structure of the IR P-L relations; the slopes are shallower.  The second reason for the difference is that the period range of the samples.The Cepheids used by \cite{Freedman2008}  and \cite{Madore2008} have periods ranging from about 6 days to 50 days, with 83\% of Cepheids having periods greater than 10 days.  The sample of Cepheids used by \cite{Ngeow2008b} and in this paper are mostly shorter period. To test the sensitivity of the slope to the periods, we computed the IRAC P-L relations for our dataset using different minimum periods, after removing those Cepheids with angular  separation $> 0.5''$ between the surveys and also removing those Cepheids that are consistent with mass loss.  We find that when the shorter period Cepheids are removed from the sample, the slopes of the IRAC P-L relations steepen. For a minimum period of $\log P = 1.05$, the slopes of the IRAC P-L relations, computed using a $3 \sigma$ iterative method, are consistent with the results of 
\cite{Freedman2008} and \cite{Madore2008}; those P-L relations 
are also listed in Table \ref{t4}. 

\section{Consistency of the Different Data Sets}
 It is important to note that the sample of Cepheids used by \cite{Freedman2008} and \cite{Madore2008}  are consistent with the sample used here as concerns the hypothesis of Cepheid mass loss.  In the  \cite{Freedman2008} and \cite{Madore2008}  samples there are no Cepheids with significant IR excesses. This is also true of the sample used here.  Of the 200 Cepheids with $P > 10$ days having  $\chi^2 < 1.25$ in the average SAGE epoch that we analyzed for the statistical significance of mass loss, we find only two Cepheids that are consistent with mass loss.  From the consistency of our sample and the sample used by  \cite{Freedman2008} and \cite{Madore2008}, and from the inherent difficulty of determining mass loss from the data of long period Cepheids, we suggest that these samples are all affected by mass loss to the same degree.  Taking period cuts of the predicted IR fluxes and computing the best-fit IR P-L relations predicts slopes that become shallower with increasing period. The observed steeper P-L relations defined by the sample of longer period Cepheids is consistent with the predictions of \cite{Neilson2008b} that mass loss for these Cepheids is driven primarily by radiation, without major contributions from shocks and pulsation, with the mass-loss rates ranging from $10^{-9}- 10^{-8} M_{\odot}/yr$. Radiative-driven mass loss is period dependent, which leads to an infrared excess compared to the stellar luminosity for Cepheids with $P > 10$ days. Therefore mass loss might cause the steeper IR P-L relations. 
\begin{figure*}[t]
\begin{center}
	\epsscale{1.15}
		\plottwo{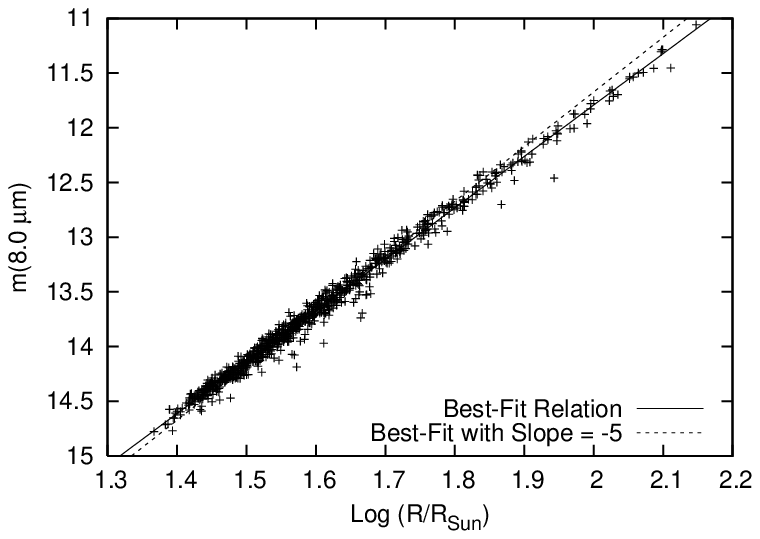}{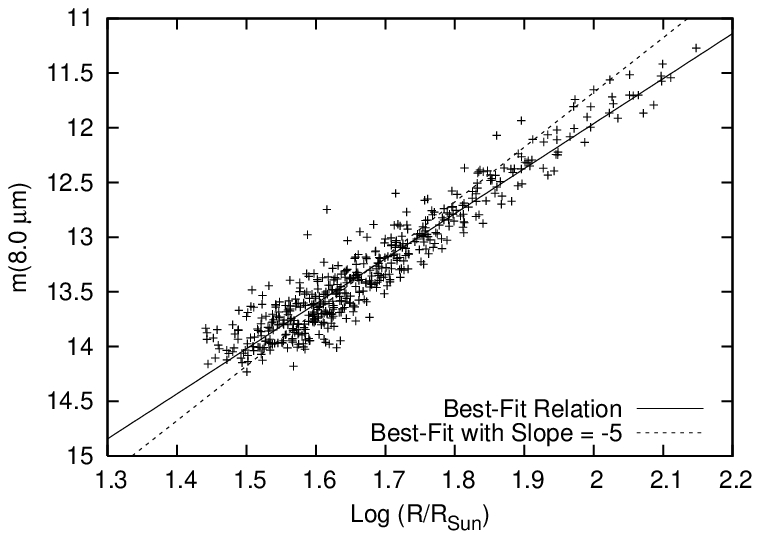}
\caption{The (Left) predicted and (Right) observed $8.0~\mu m$ flux of the LMC Cepheids as a function of the predicted radii of the Cepheids.  The solid line is the linear least squares best-fit to the data, while the dotted line is the linear least squares best-fit with the slope forced to be $-5$, which is the expected slope if the temperature of the star does not effect the flux. }
	\label{f10}
	\end{center}
\end{figure*}

Another consistency check is the correspondence between the P-R and the P-L relations.  \cite{Freedman2008} were able to derive the slope of the P-L relation in the limit of long wavelengths by using the result from \cite{Gieren1999} that the slope of the P-R relation is about 0.68.  Because the long wavelength flux is primarily dependent on the Cepheid's radius, they ware able to find that the slope of the P-L relations is about $-3.4$. In this work, we find that the slope of the P-R relation is the same as that of \cite{Gieren1999} but the slope of the predicted IR P-L relations range from -3.3 to -3.2 significantly different from that asymptotic limit.  We investigate the apparent contradiction by re-deriving the asymptotic limit to the slope of the P-L relation and plotting the predicted and observed $8.0$ $\mu m$ fluxes as a function of the predicted radius in Figure \ref{f10}.   In Equation \ref{e1} we had that $L_\nu \propto R^2B_\nu(T_{\rm{eff}})$, which means the flux is $f_\nu  \propto R^2B_\nu(T_{\rm{eff}})/d^2$. The distance $d$ is that of the LMC and is assumed to be the same for all Cepheids in our sample.  Converting the flux to apparent magnitude yields
\begin{equation}\label{e8}
m_\lambda = -5\log R - 2.5 \log B_\nu(T_{\rm{eff}}) + C_1,
\end{equation}
where $C_1$ is a constant and contains the distance.  If the slope of the linear best fit to the magnitude versus the $\log R$ is $-5$ then there is a serious problem with our analysis as it suggests that the Blackbody function does not contribute to the brightness in the infrared.  However, the best-fit relations have slopes of $-4.70$ and $-4.12$ for the predicted and observed fluxes respectively.  This suggests that the radius of the Cepheid is related to the Blackbody radiation via the effective temperature and pulsation period.  The difference between the slopes  corresponds to the slopes of the respective P-L relations: $-3.49$ from the \cite{Madore2008} sample, $-3.28$ from our predicted fluxes, and $-2.85$ from the observed fluxes.  As expected the flux-radius relations do not have a slope of $-5$ because the effective temperature affects the fit of both the predicted and observed fluxes, even at long wavelengths via its relation to the radius. Using the observed fluxes, the apparent IR excesses cause the slope to become even shallower. We show this by replacing the blackbody function at infrared wavelengths by the Rayleigh-Jeans approximation, $B_{\nu}(T_{\mathrm{eff}}) \propto T_{\mathrm{eff}}$, giving 
\begin{equation}\label{e9}
m_\lambda = -5\log R -2.5 \log T_{\rm{eff}} + C_2.
\end{equation}
\cite{Beaulieu2001} have shown that the effective temperature is a function of the gravity and $(V-I)_0$ color, $ \log T_{\rm{eff}} = 3.92 +0.0055\log g -0.2487(V-I)_0$. The color term can be replaced by the period via a Period-Color (P-C) relation and the gravity term is $\log g = 2.62-1.21\log P$.  For instance, using the P-C relation from \cite{Sandage2004} for the LMC Cepheids, $(V-I)_0 = 0.256\log P + 0.444$, gives
\begin{equation}\label{e10}
m_\lambda = -5 \log R + 0.175 \log P + C_3.
\end{equation}
If we use the P-R relation to replace the period term, we can derive the limiting slope to the  flux-radius relation, 
\begin{equation}\label{e11}
m_\lambda = -4.74\log R + \rm{Constant}.
\end{equation}
Or we can use the P-R relation to replace the radius term to find the limiting slope of the P-L relation,
\begin{equation}\label{e12}
m_\lambda = -3.23\log P + \rm{Constant}.
\end{equation}
The slope of Equation \ref{e11} agrees with the best-fit slope of the flux-radius relation from Figure \ref{f10}, suggesting that indeed the effective temperature affects the slope of the P-L relation at long wavelengths.  Therefore the asymptotic limit of the slope of the P-L relation at long wavelengths is approximately $-3.23$.  However, this does not necessarily mean that the P-L relations of \cite{Freedman2008} and \cite{Madore2008} are incorrect. If the slope of the P-R relation is $0.71$, the asymptotic limit of the slope of the P-L relation is $-3.4$, still consistent with observed P-R relations found by \cite{Gieren1998} and  \cite{Laney1995}.  

\section{Predictions of the Non-Linear Period-Luminosity Relations}
\cite{Ngeow2008a} and \cite{Ngeow2008b} found the surprising result that at 5.8 and at 8.0 $\mu$m the slopes of the P-L relations become steeper for Cepheids with $P > 10$ days.  We want to test if this could be caused by mass loss that makes the predicted P-L relations non-linear.  To do this, we have used two linear P-L relations, one for periods less than 10 days and the other for longer periods.  These are of the form 
\begin{equation}\label{e13}
m_\lambda =\left\{ \begin{array}{l} a \log P + b \mbox{\hspace{1cm}} \log P < 1 \\
c \log P + d\mbox{\hspace{1cm}} \log P >1 \end{array}\right. .
\end{equation}

We compute the best-fit non-linear relations employing the same $3\sigma$-iterative routine used before. The $F$-test is used to check if the non-linear relations are statistically a better fit to the data and the relations are listed in Table \ref{t5}.  We find that {\it{all}} P-L relations at {\it{all}} wavelengths for each epoch have $F > 3$, meaning they are consistent with being non-linear.  Mass loss affects the structure of the P-L relations and the potential value of $F$ computed for the IR P-L relations.  This indicates that the presence of mass loss acts to hide the non-linear nature of the IR P-L relations, steepening the slopes for $\log P >1$ of the observed $5.8$ and $8.0$ $\mu m$ P-L relations. All  the predicted P-L relations have slopes for $\log P >1$ that are shallower than the slopes for $\log P <1$ for the same wavelength for second and average epoch fluxes. The first epoch fluxes have non-linear P-L relations where the slopes for the long-period Cepheids is approximately the slope of the P-L relation for the short-period Cepheids within the uncertainty.  Because the errors of the slope of the long-period relations is large then the structure of the first epoch non-linear P-L relations are still consistent with the structure of the non-linear P-L relations for the other epochs.  This also shows why the slopes of the predicted IR linear P-L relations increases from $-3.3$ to $-3.2$ when the short-period Cepheids are removed from the sample, while the opposite effect seen for the observed Cepheid fluxes.
\begin{table*}[t]
\caption{Best Fit Parameters for Predicted Non-Linear Period-Luminosity Relations}
\begin{center}
\begin{tabular}{ccccccccc}
\hline
Band & Slope$_S$ & Zero Point$_S$ &  Slope$_L$ & Zero Point$_L$ & Dispersion & N & F \\
\hline
\multicolumn{8}{c}{Epoch 1} \\
\hline
V& $-2.917\pm 0.056$&$17.344\pm 0.037$&$-2.457\pm 0.273$&$16.958\pm 0.313$&$0.202$&$707$&$7.09$\\
I& $-3.068\pm 0.041$&$16.507\pm 0.027$&$-2.743\pm 0.196$&$16.246\pm 0.225$&$0.146$&$707$&$8.59$\\
J&$-3.185\pm 0.030$&$ 16.366\pm 0.020$&$ -2.932\pm 0.139$&$ 16.155\pm 0.160$&$0.108$&$708$&$7.87$\\
H&$-3.218\pm 0.027$&$ 16.127\pm 0.018$&$-3.113\pm 0.125$&$ 16.073\pm 0.143$&$ 0.098$&$709 $&$ 6.33$\\
K&$-3.255\pm 0.025$&$ 16.042\pm 0.016$&$-3.162\pm 0.113$&$15.995\pm 0.130$&$0.089$&$710$&$5.99$\\
$3.6$ & $-3.296\pm 0.022$&$16.006 \pm 0.015$&$-3.281\pm 0.101$&$16.033\pm 0.116$&$0.078$&$705$&$4.59$\\
$4.5$& $-3.310\pm 0.021$&$15.933 \pm 0.014$&$-3.304\pm 0.100$&$ 15.970\pm 0.114$&$0.078$&$706$&$4.39$\\
$5.8$&$-3.321\pm 0.021$&$15.936\pm 0.014$&$ -3.325\pm 0.099$&$15.981\pm 0.113$&$0.077$&$707$&$4.13$\\
$8.0$&$-3.327\pm 0.021$&$15.936\pm 0.014$&$-3.343\pm 0.097$&$ 15.993\pm 0.111$&$0.076$&$706$&$3.68$\\
 \hline
\multicolumn{8}{c}{Epoch 2} \\
\hline
V&$-2.928\pm 0.053$&$ 17.360\pm 0.035$&$ -2.501\pm 0.208$&$17.005\pm 0.241$&$ 0.195$&$713$&$7.73$\\
I&$-3.077\pm 0.039$&$ 16.519\pm 0.26$&$ -2.767\pm 0.152$&$ 16.263\pm 0.176$&$ 0.143$&$716$&$8.09$\\
J&$-3.190\pm 0.029$&$16.372\pm 0.019$&$-2.945\pm 0.111$&$16.165\pm 0.129$&$ 0.107$&$718$&$8.08$\\ 
H&$-3.219\pm 0.027$&$16.129\pm 0.18$&$-3.081\pm 0.101$&$16.032\pm 0.116$&$ 0.098$&$722$&$ 5.94$\\
K&$-3.255\pm 0.024$&$16.043\pm 0.016$&$ -3.147\pm 0.092$&$ 15.968\pm 0.107$&$0.090 $&$724$&$4.66$\\
$3.6$&$-3.299\pm 0.021$&$16.006\pm 0.014$&$-3.214\pm 0.081$&$15.954\pm 0.094$&$ 0.079$&$ 715$&$ 4.60$\\
$4.5$& $-3.311\pm 0.021$&$15.933\pm 0.014$&$-3.197\pm 0.081$&$15.848\pm 0.093$&$0.077$&$715$&$ 5.37$\\
$5.8$&$-3.320\pm 0.021$&$ 15.933\pm 0.014$&$-3.213\pm 0.080$&$ 15.853\pm 0.092$&$ 0.076$&$715$&$4.85$\\
$8.0$&$-3.327\pm 0.021$&$15.934\pm 0.014$&$-3.227\pm 0.079$&$ 15.861\pm 0.091$&$ 0.076$&$ 715$&$ 4.36$\\
\hline
\multicolumn{8}{c}{Average of Two Epochs} \\
\hline
V&$-2.882\pm 0.050$&$17.329\pm 0.033$&$-2.429\pm 0.210$&$16.921\pm 0.245$&$0.198$& $826$&$6.56$\\
I&$-3.045\pm 0.036$&$16.498\pm0.024$&$-2.775\pm 0.153$&$ 16.273\pm 0.177$&$ 0.144$&$828$&$ 6.45$\\ 
J&$-3.170\pm 0.028$&$ 16.361\pm 0.018$&$-2.921\pm 0.114$&$16.137\pm 0.133$&$0.110$&$ 835 $&$ 6.72$\\
H&$-3.200\pm 0.025$&$16.118\pm 0.016$&$ -3.066\pm 0.100$&$ 16.016\pm 0.116$&$ 0.098$&$ 835$&$ 4.98$\\
K&$-3.238\pm 0.022$&$ 16.033\pm 0.015$&$ -3.121\pm 0.091$&$ 15.944\pm 0.106$&$ 0.089$&$836 $&$4.66$\\
$3.6$&$-3.284\pm 0.020$&$15.999\pm 0.013$&$ -3.211\pm 0.080$&$ 15.952\pm 0.093$&$0.078$&$828$&$ 3.61$\\
$4.5$&$-3.290\pm 0.019$&$15.921\pm 0.013$&$ -3.200\pm 0.078$&$15.852\pm 0.091$&$ 0.076$&$ 826$&$3.53$\\ 
$5.8$&$-3.309\pm 0.019$&$ 15.928\pm 0.012$&$ -3.217\pm 0.078$&$ 15.860\pm 0.091$&$ 0.076$&$ 829$&$4.03$\\
$8.0$&$-3.316\pm 0.019$&$15.930\pm 0.012$&$-3.233\pm 0.077$&$15.869\pm 0.090$&$ 0.075$&$829$&$3.60$\\
\hline 
\end{tabular}
\end{center}
\label{t5}
\end{table*}

In section 9 we based our analysis on the asymptotic limit of the slope of the \emph{linear} IR P-L relations.  Here we want to use this limiting behavior to test if it is plausible that the IR P-L relations are actually non-linear. In section 9 we used Equation \ref{e1} and the Rayleigh-Jeans approximation for the blackbody function, which enabled use to write the effective temperature as a 
function of the $(V-I)_0$ color and then as a function of period via the Period-Color relation. Here we replace the linear Period-Color relation with the non-linear Period-Color relation from \cite{Sandage2004}, which has a slope of 0.160 for $\log P < 1$ and a slope of 0.315 for $\log P> 1$.  Doing this readily shows that the IR P-L relation is non-linear at long wavelengths. Furthermore, the P-L relation is the sum of a linear P-R relation and a non-linear Period-Color relation. In that case the limiting slope of the non-linear P-L relation is -3.30 for $\log P < 1$ and the limiting slope is -3.21 for $\log P > 1$. These values are consistent with the non-linear IR P-L relations we have derived here,  but we note that the analysis is very sensitive to the slope of the  P-R and Period-Color relations. An increase of only 0.01 to the slope of the P-R relation causes the P-L relations to be steeper by 0.05.  Also, \cite{Kanbur2004} found a non-linear Period-Color relation for LMC Cepheids with a slope of 0.152 for $\log P < 1$ and a slope of 0.590 for $\log P > 1$. The limiting slopes of the non-linear IR P-L relations is then 3.31 and 3.04. 

Our derivation also provides insight into why \cite{Ngeow2008b} found the IR P-L relations that are consistent with  linearity.  At any wavelength the P-L relation is the combination of the P-R relation and the Period-Color relation, connected via the effective temperature.  The P-R relation has a slope and zero-point that are independent of wavelength, but the effective temperature contributes to the luminosity through the wavelength-dependent blackbody function. At wavelengths in the Rayleigh-Jean regime, the blackbody function is a linear function of the effective temperature, while at optical wavelengths the greater dependence of the full Planck  function on effective temperature leads to a more dramatic change in luminosity, i.e. the slope of non-linear Period-Color relation has a much more important contribution to the structure P-L relation.  At longer wavelengths, the non-linear Period-Color relation plays a much reduced role, and the change of the slope with period is minimum.  Thus, to be able to determine if the IR P-L relations are non-linear the scatter of the fit of observed fluxes of Cepheids must be significantly smaller and have any contamination of IR excess from mass loss removed.
\begin{figure*}[t]
\begin{center}
	\epsscale{1.15}
		\plottwo{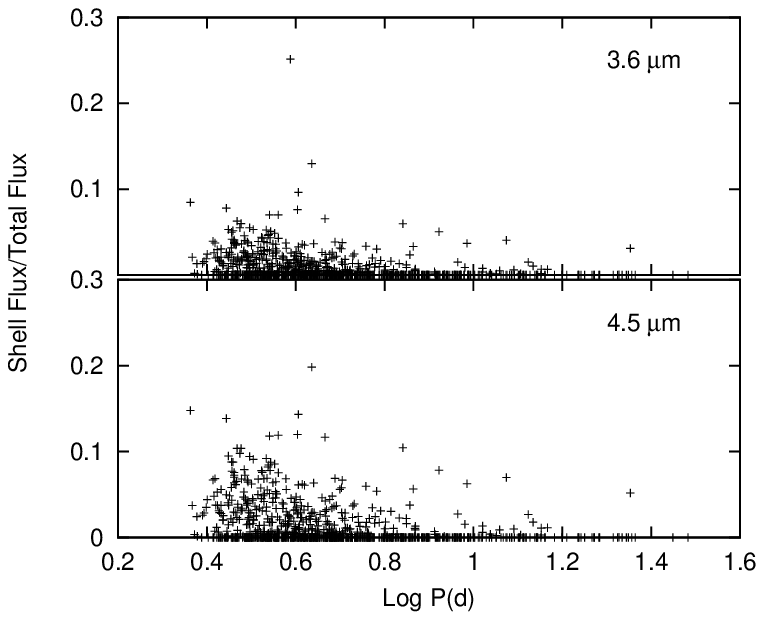}{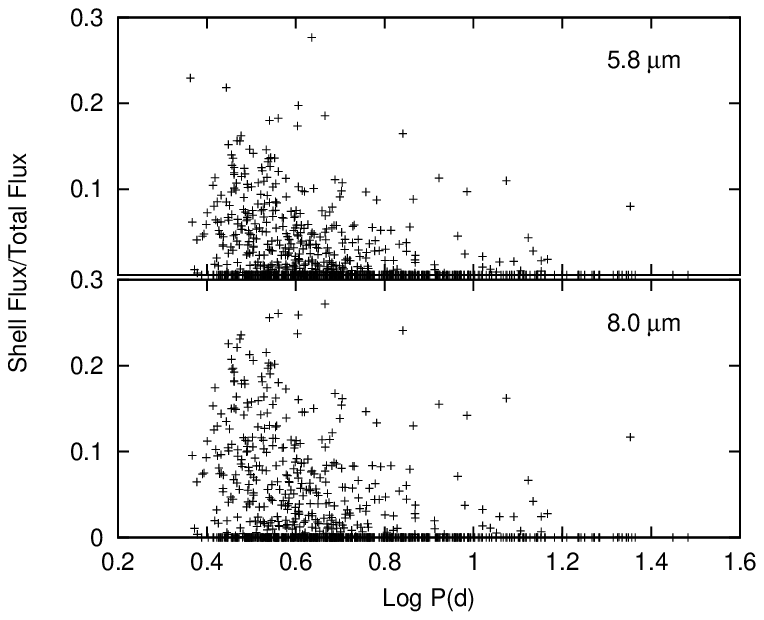}
\caption{The predicted ratio of the luminosity of the dust shell surrounding each Cepheid generated by mass loss to the total (shell + stellar) luminosity of each Cepheid in the four IRAC wavelengths as a function of pulsation period.}
	\label{f11}
	\end{center}
\end{figure*}

Our predicted linear P-L relations are similar to the observed relations from \cite{Ngeow2008a} and \cite{Ngeow2008b}, suggesting that mass loss does not contribute much to the IR observations.  However, the non-linear IR relations are significantly different.   For $\log P > 1$ the relations have predicted slopes that are shallower than the observed relations, while for $\log P < 1$ the predicted slopes are steeper than the slopes of the observed relations. These results suggest that mass loss may play two roles, as can be seen in Figure \ref{f11}, which plots the ratio of the flux from the dust shell to the total flux of the Cepheids as a function of the pulsation period for each IRAC band. The first role of mass loss is to increase the dispersion of the P-L  relations at all period cuts, and the second role is to alter the slope of the linear and non-linear P-L relations. 

The dispersion of the IR P-L relations for LMC Cepheids is caused by several factors: the width of the Cepheid instability strip, the tilt and depth of the LMC and possibly mass loss. Considering the geometry and the metallicity distribution of the LMC,  \cite{Madore2008} derived the minimum observable dispersion of the linear IR P-L relations to be about 0.08 mag. Our analysis here finds the dispersions of the linear IR relations to range from 0.079 mag at 3.6 $\mu$m to 0.074 mag at 8.0 $\mu$m.  These  dispersions are consistent with the minimum dispersion found by  \cite{Madore2008}, which suggests that mass loss contributes only 0.001 to 0.006 mag to the dispersion of the IR P-L relations.

The behavior of mass loss in LMC Cepheids affects the structure of  the IR P-L relations. As shown in Figure \ref{f11}, for shorter-period Cepheids ($\log P < 1$) the shell flux is a larger fraction of the total flux, while for longer-periods Cepheids the ratio is approximately constant within some dispersion.  The large number of short-period Cepheids in the sample causes the slopes of the linear IR P-L relations to be shallower, and as wavelength increases, these slopes become even more shallow. This behavior explains why the slopes of the IR P-L relations determined by \cite{Ngeow2008b} become shallower as a function of increasing wavelength. Mass loss likewise affects the structure of the non-linear IR P-L relations.  For the short period part of the non-linear IR P-L relations, mass loss causes the slope to be shallower, while for the long period Cepheids, the slope is steepened.  At $3.6$ $\mu m$ and $4.5$ $\mu m$,  this, along with an increased dispersion, makes the P-L relations appear to be consistent with linearity by causing a smaller computed value of $F$ for the $F$-test.   At longer wavelengths this effect is more pronounced; mass loss causes the slope of the P-L relations for $\log P <1$ to be more shallow than the long-period relations, opposite of what has been determined for the optical and near-IR relations.   

The predicted linear and non-linear P-L relations, as well as the effects of mass loss on their structure, agree with the results of our previous work using SAGE epoch1-data correlated with the OGLE-II Cepheids.  In that work the IR P-L relations were found to be shallower than those found here, the difference being due to the use of a stricter $\chi^2$ cutoff, which we noted has an effect on the predicted slope of the P-L relations.  Regardless of this difference, we have shown in both works that the IR P-L relations are affected by infrared excess caused by the presence of dust shells similar to those that have been observed \citep{Kervella2006, Merand2006, Merand2007, Marengo2009}.   Because the amount of circumstellar dust is dependent on the dust-to-gas ratio, the infrared excess is metallicity dependent.  The IR P-L relations, therefore, are metallicity dependent as shown in the previous work.

\section{Conclusions} 
The purpose of this work is to use the OGLE-III sample of fundamental-mode Cepheids, complemented by data from 2MASS and two epochs of SAGE observations from \cite{Ngeow2008b} to test for the existence of infrared excess due to the formation of dust shells at a large distance from the surface of a Cepheid.  The dust shells are hypothesized to be caused by mass loss from the Cepheids.  In addition to testing for the existence of IR excess, we have investigated how the IR excess affects the structure of the P-L relations, whether the P-L relations are consistent with being non-linear and the uncertainty of the infrared surface brightness technique due to mass loss.

We fit two models to  the $VIJHK$ and IRAC observations; the first model fits the observed fluxes of  Cepheids using just one free parameter, the radius of the Cepheids.  The second model uses two free parameters, the radius and the dust mass-loss rate which is converted to a gas mass-loss rate.  In both cases, the effective temperature is calculated using the relation from \cite{Beaulieu2001}.    The fits of the two models are compared using the $F$-test,  and we found that the sample is consistent with mass loss.   In addition to having large mass-loss rates, these Cepheids tend to  have shorter periods, $\log P< 1$. This is because long-period  Cepheids are cooler and have larger IR stellar fluxes, which makes  the contribution of a circumstellar dust shell a smaller relative contribution or IR excess.  In this work, we also use a stricter  $\chi^2$ cutoff than in our previous paper, which leads us to predict $V$- and $I$-band P-L relations that agree with observations.  The use of the stricter cutoff removes Cepheids that are poor fits due to being at a distance very different from the assumed LMC distance modulus of $18.5$ and having IR fluxes that are significantly different from the mean brightness.

Because mass loss causes an infrared excess, it also affects the IR surface brightness technique for determining the angular diameter of Cepheids by increasing the observed $J$- and $K$-band fluxes to be brighter than the stars brightness.  For the $V$- and $K$-band relation, the value of $(V-K)_0$ is larger and the computed angular diameter is overestimated.  The error of the relation is about $3.5\%$. Using the $J$- and $K$-band relation, there are offsetting errors because it uses the color $(J-K)_0$ and the $K$-band brightness, and has an uncertainty of about $0.4\%$.  Also, it is found that the calibration of the IRSB method has error due to the possibility of mass loss in the calibrating Cepheids.  The uncertainty due to calibration is about $2.2\%$ and $0.9\%$ for the $V, K$-band and $J, K$-band relations, respectively. 

 Using our predicted radii, we calculated the Period-Radius relations for each epoch of SAGE observations. We find that the P-R relations have a slope of $0.690 \pm 0.003$ and a zero-point of $1.128 \pm 0.002$.  The slope and zero-point agree with the slope and zero-point found by \cite{Gieren1999}; however the zero-point in our relation is notably smaller.

We predicted the linear P-L relations for the LMC Cepheids, finding that the IRAC P-L relations have slopes that range from $-3.22$ to $-3.26$.  These slopes agree with the results of \cite{Ngeow2008b} within the dispersion, but not with the results of \cite{Freedman2008} or \cite{Madore2008}.  Because the $F$-test predicts that about half of the Cepheids are consistent with no mass loss, we compute the IR P-L relations for the sample of Cepheids after removing those that appear to have mass loss as indicated by an infrared excess.  These relations are computed without the period cut-offs suggested by \cite{Ngeow2008b}, and it is clear the relations are not significantly affected by the removal of the Cepheids that might have mass loss.  This suggests that, although the IR excess might still important in the sample, it cannot be statistically identified because the uncertainty of the observations is so large.   The non-linear P-L relations are calculated and tested using the $F$-test, and the P-L relations are non-linear at all wavelengths.  This result differs from the observations where the relations are found to be non-linear at wavelengths shorter than $K$-band.  

We found that the IR excess due to mass loss is strongly influenced by the unknown phase of the IRAC fluxes.  For instance, if the amplitude of light variation for a Cepheid in the IR is $0.1$ magnitude then the ratio of the maximum luminosity to the mean luminosity $L_{max}/L_{mean} \approx 1.1$ This would appear to be equivalent to an IR excess from a shell $L_{shell}/L_{mean} \approx 10\%$.  A comparison to the shell fluxes shown in Figure \ref{f11} suggests that the unknown pulsation phase of the IR fluxes is the most important uncertainty in our analysis.  Time-series observations of Cepheids in the IR with complete phase coverage is necessary to better understand how mass loss works in Cepheids and how the IR excess affects the IR P-L relation and IRSB technique.

All of these results suggest that mass loss in Cepheids has an important effect on the structure of the IR Period-Luminosity relations and the Period-Radius relations that are predicted from the IR surface brightness technique. We have shown that mass loss occurs in half of the sample of OGLE-III Cepheids, making it difficult to state conclusively that mass loss is an important phenomena in all Cepheids. If the uncertainty of the fluxes were less, more Cepheids might be consistent with having a stellar wind. While having a larger sample of Cepheids and two epochs of SAGE observations has provided a better testbed for studying mass loss, even more data are needed. Fitting mass loss and radii of Cepheids is better done with time series observations of the Cepheids to determine the mean IR flux and decrease the phase uncertainty. The uncertainty of the IR brightness of Cepheids due to mass loss needs to be characterized if the IR P-L relations are to be used as tools for measuring precise distances to galaxies and determining the Hubble constant to an accuracy of 2-3\%. This issue may only be resolved by more observations of Cepheids in IR wavelengths. 

\begin{acknowledgements}
HN would like to thank Dr Tom Barnes, and Dr Jesper Storm for enlightening discussions about IRSB technique.  We would also like to thank the anonymous referee for helpful comments. 
\end{acknowledgements} 

\bibliography{wind_th}
\bibliographystyle{apj}

\end{document}